\documentclass[aps,prb,amsmath,twocolumn,amssymb,floatfixng,showpacs,
superscriptaddress,footinbib]{revtex4-1}
\pdfoutput=1
\usepackage[dvips]{graphics}
\usepackage{hyperref}
\usepackage{bm}
\usepackage{epsfig}
\usepackage{enumerate}
\usepackage{color}
\usepackage{graphicx}
\definecolor{dred}{rgb}{0.6,0,0}
\hypersetup{
    pdfnewwindow=true,      % links in new window
    colorlinks=true,       % false: boxed links; true: colored links
    linkcolor=magenta,          % color of internal links
    citecolor=blue,        % color of links to bibliography
    filecolor=blue,      % color of file links
    urlcolor=blue        % color of external links
}

\newcommand\bea{\begin{eqnarray}}
\newcommand\eea{\end{eqnarray}}
\newcommand\beq{\begin{equation}}  
\newcommand\eeq{\end{equation}}
\newcommand{\non}{\nonumber} 
 
\newcommand{\ie}{{\it i.e. }}

\newcommand{\etal}{{\it et al. }}
 
\newcommand\mb{\mathbf}
\begin{document}

\title{Aharonov-Bohm effect in a helical ring with long-range hopping: Effects of Rashba spin-orbit 
interaction and disorder}

\author{Paramita Dutta}
\email{paramitad@iopb.res.in}
\affiliation{Institute of Physics, Sachivalaya Marg, Bhubaneswar-751005, India} 
\author{Arijit Saha}
\email{arijit@iopb.res.in}
\affiliation{Institute of Physics, Sachivalaya Marg, Bhubaneswar-751005, India} 
\author{A. M. Jayannavar}
\email{jayan@iopb.res.in}
\affiliation{Institute of Physics, Sachivalaya Marg, Bhubaneswar-751005, India} 

\begin{abstract}
We study Aharonov-Bohm effect in a two-terminal helical ring with long-range hopping in presence of Rashba 
spin-orbit interaction. We explore how the spin polarization behavior changes depending on the applied 
magnetic flux and the incoming electron energy. The most interesting feature that we articulate in this system is that 
zero-energy crossings appear in the energy spectra at $\Phi=0$ and also at integer multiples of half-flux 
quantum values ($n \Phi_0/2$, $n$ being an integer) of the applied magnetic flux. We investigate the transport 
properties of the ring using Green's function formalism and find that the zero energy transmission peaks 
corresponding to those zero energy crossings vanish in presence of Rashba spin-orbit interaction. We also 
incorporate static random disorder in our system and show that the zero energy crossings and transmission 
peaks are not immune to disorder even in absence of Rashba spin-orbit interaction. The latter prevents the 
possibility of behaving these helical states in the ring like topological insulator edge states.  
 
\end{abstract}

\pacs{73.23.-b,72.25.-b,71.70.Ej}

\maketitle

%------------------------
\section{Introduction}
%-----------------------

During the last two decades manipulating and controlling electronic spin, one of the most fundamental degrees of 
freedom, has introduced a new paradigm in the field of quantum condensed matter physics especially in 
spintronics~\cite{wolf2001spintronics,vzutic2004spintronics}. It has drawn interests of the scientific community 
due to the prospects of application in modern quantum devices. Till date, various spintronic phenomena like 
spin-switching, spin filtering, spin-splittering etc. have been proposed using ferromagnetic leads, external 
magnetic field and so on~\cite{frustaglia2001quantum,koga2002spin}. Internal properties 
of the system have also been utilized for maneuvering the spin current in a precise way~\cite{lu2007pure}. Internal 
properties like spin-orbit interaction (SOI), especially Rashba spin-orbit interaction 
(RSOI)~\cite{bychkov1984oscillatory,bercioux2015quantum} which arises due to the structural inversion asymmetry 
yields an alternative way of spin manipulation~\cite{egues2002rashba}. The tunability of RSOI strength by external 
gate voltage placed in vicinity of the sample offers an additional degree of freedom in this context~\cite{nitta1997gate}.

A large number of spintronic devices have been proposed on the basis of quantum interferometers among 
which the simplest one is the ring geometry~\cite{wang2008anisotropic}. The quasi-one-dimensional Aharonov-Bohm 
rings are elegant test-beds for exploring the quantum coherence 
phenomena~\cite{cheung1988isolated,buttiker1984quantum,cheung1988persistent,yuvalgefen2,deo1994quantum,jayannavar1994persistent,jayannavar1995persistent}. 
The simplicity of its topological geometry has drawn attention of researchers due to their potential application in 
various nano-electronic devices. Research activity on these ring structures have been boosted after the recent 
advancement of nano-technology which has made it possible to fabricate metallic as well as semi-conductor rings in desired 
way~\cite{deblock2002diamagnetic,mclellan2004edge,sun2006construction,amiryacoby,glazmanvonoppen}. To study the 
interference effect in such a ring geometry, Aharonov-Bohm (AB) flux is a key ingredient as it affects only the 
phases of the electronic wave functions~\cite{aharonov1959significance}. In ordinary mesoscopic ring AB effect deals 
with the charge of the electrons keeping the spin degeneracy intact. Therefore, breaking the spin degeneracy can be 
an interesting aspect~\cite{moldoveanu2010tunable}. This invokes one to think about the spin AB effect where the 
spin manipulation can be done by AB flux utilizing the spin-dependent phase factors introduced by SOI. 
This is a non-local phenomenon~\cite{maciejko2010spin}. 

Indeed, this type of spin manipulation by magnetic flux can be done by using the helical edge states which is 
observed in topologically non-trivial new state of matter~\cite{rod2015spin,jiang2009quantum}. At the boundary 
of a two-dimensional ($2$D) topological insulator (TI), a pair of counter-propagating states with two opposite 
spins and protected by time-reversal symmetry appear~\cite{hasan2010colloquium,sczhangreview}. These $1$D gapless helical 
edge states are robust to static disorder resulting in the absence of back-scattering, and they cannot be perturbed
as long as time-reversal symmetry is preserved. They were theoretically predicted in quantum spin Hall 
insulators~\cite{kanemele2005,qi2010quantum}. Later, Bernevig \etal\cite{bernevig2006quantum} have investigated the band 
structures of HgTe/CdTe heterostructrues and shown a transition from topologically trivial to non-trivial 
state of matter by means of change in width of the heterostructure. However, the existence of these $1$D edge 
states was experimentally realized in HgTe/CdTe heterostructrues by K\"{o}nig \etal\cite{konig2007quantum}. 
This path-breaking observation stimulated the research in this direction both 
theoretically~\cite{culcer2010two,sun2011nearly,strom2010edge,strom2015controllable} and 
experimentally~\cite{moore2009topological,zhang2009experimental}. In order to understand the behavior of these 
helical edge states characterized by linear dispersion, various models have been proposed in literature like, 
Kane Mele model~\cite{kanemele2005,kane2005z}, Bernevig-Hughes-Zhang (BHZ) model~\cite{bernevig2006quantum} etc. 
Parallel to the continuum model, lattice models have also been proposed to describe this state of matter being 
topologically distinct from all other known states of matter~\cite{jiang2009numerical}. Although it is difficult 
to map them in a lattice model since one faces the problem of Fermion doubling~\cite{nielsen1981no}. 

In a very recent paper Masuda \etal have suggested a model where one can get rid of the Fermion doubling 
problem~\cite{masuda2012interference} in a lattice model. They have considered a $1$D ring with long-range hopping 
that stimulates the $1$D edge states of a $2$D topological insulator neglecting the lateral distribution of wave function 
in real system. This long-range hopping model was actually introduced by Gebhard and Ruckenstein~\cite{gebhard1992exact}. 
Nevertheless, their purpose was to explore Mott-Hubbard metal-insulator transition in this model. While, Masuda 
\etal have modified their model to obtain a $1$D interferometer with helical spin current and predicted a way 
to generate spin polarized current by injecting completely spin unpolarized electrons. 

This type of $1$D model with long-range hopping have already been used earlier to describe several physical systems like, 
cold atoms, ion traps 
etc~\cite{jurcevic2014observation,richerme2014non,cheraghchi2005localization,zhang2002localization,xiong2003scaling,loring1984hopping,della1993energy,de1998delocalization,levitov1999critical} also in 
bio-molecules~\cite{yamada2004electronic,giese2000long,schuster2000long}. Very 
recently, Celardo \etal have studied shielding and localization phenomena in a paradigmatic model of $1$D ring 
with long-range hopping~\cite{celardo2016shielding}. Whereas, manifestation of $1$D edge states by a simple ring 
geometry is a new one. However, Masuda \etal~\cite{masuda2012interference} have not considered the role of Rashba
spin-orbit interaction which plays a crucial role in modeling the quantum spin Hall states. In addition to this, 
the phenomenon of quantum interference can be more interesting in presence of AB flux. Now, the effect of AB flux in 
$2$D TI ring is already well-explored phenomenon~\cite{peng2010aharonov,michetti2011bound}. Most of the previous works 
were done in order to study the circulating current within the closed boundary. A very few studies have been made to 
investigate the transmission or the conductance of electrons in open systems comprising of $1$D helical states 
being attached to leads in presence of magnetic flux. As an example, Chu \etal have shown periodic 
oscillations in magnetoconductance leading to a possibility of giant magnetoresistance that may be utilized in 
practical application~\cite{chu2009coherent}. 

Motivated by the above mentioned facts, we explore the spin-dependent transport phenomena in a $1$D ring with 
long-range hopping in presence of AB flux and Rashba SOI. We predict the existence of zero-energy crossings at zero 
and integer multiples of half-flux quantum values of the applied magnetic flux. We attach two $1$D semi-infinite 
leads to the ring and studied the transport properties using Green's function formalism~\cite{datta1997electronic}. 
In the transmission spectra we notice the zero-energy peaks corresponding to zero-energy states in the energy 
spectra for the same flux values. However, the zero-energy crossing as well as the associated zero-energy peaks 
corresponding to the zero energy states no longer exist in presence of RSOI. Instead, a gap appears
in the energy and corresponding transmission spectra around the zero-energy value. We also incorporate 
static disorder into the system and examine whether zero-energy crossings are immune to disorder which is one of the 
essential features of $1$D helical edge states in $2$D topological insulators. We observe that the zero-energy states 
and the peaks get affected by the presence of non-magnetic impurity even in absence of RSOI. They are not robust 
to disorder. Hence, these $1$D helical states are sensitive to both RSOI and disorder.

%In what follows we arrange the rest of the manuscript. 
The remainder of this paper is organized as follows. We describe our model and Hamiltonian in Sec.~\ref{sec2} and 
the Green's function formalism in Sec.~\ref{sec3} in order to calculate the transmission probability of electrons. 
In Sec.~\ref{nr} we discuss our numerical results which include the effect of AB flux, RSOI and disorder. 
Finally, we summarize and conclude in Sec.~\ref{concl}.

%---------------------
\section{Model}
\label{sec2}
%--------------------

\begin{figure}[!thpb]
\centering
\includegraphics[width=0.90 \linewidth]{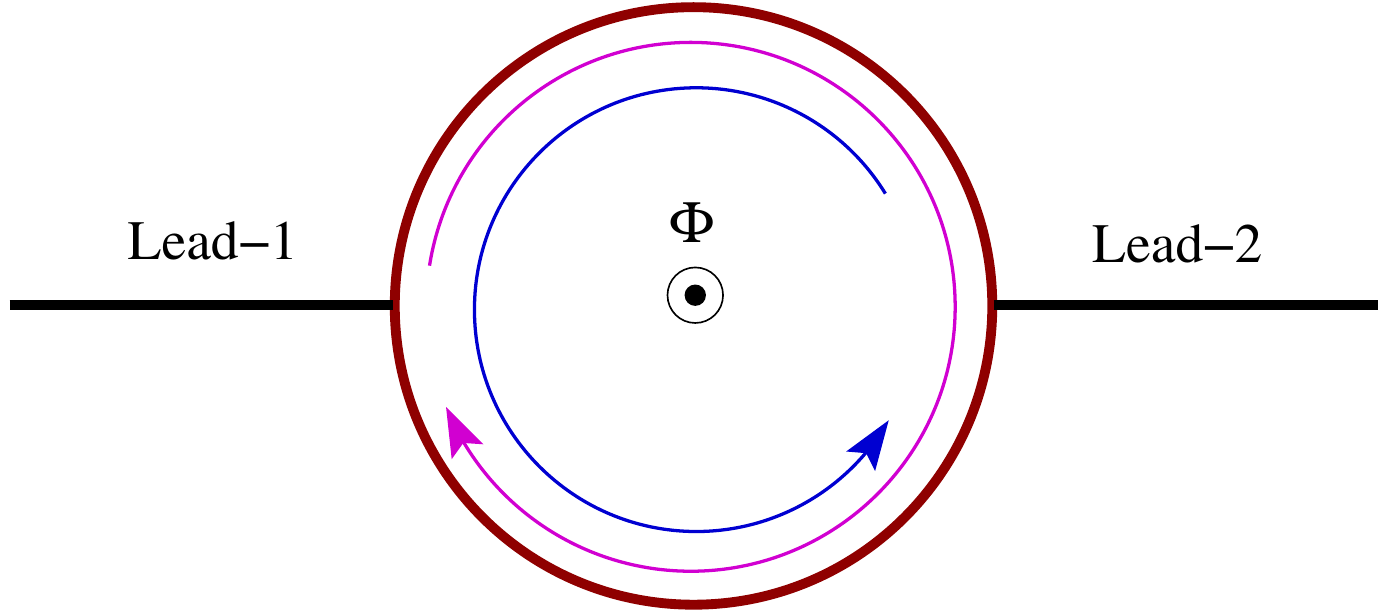}
\caption{(Color online) Schematic diagram of $1$D helical ring attached to two $1$D semi-infinite leads viz. lead-$1$
and lead-$2$. Aharonov-Bohm flux $\Phi$ is applied along the axis of the ring (upward direction). Magenta and blue 
circular arrows represent the two counter-propagating current flows with opposite spins within the ring. }
\label{model}
\end{figure}

In Fig.~\ref{model} we present our geometry in which a one-dimensional ($1$D) ring is attached to two semi-infinite 
$1$D leads namely, lead-$1$ and lead-$2$. The ring has $N$ number of atomic sites periodically arranged with the 
lattice spacing $a$. We consider $N$ as odd number and it is necessary for this model in order to get the linear 
dispersion relation as mentioned in Ref.~\onlinecite{masuda2012interference}. Two circular arrows of magenta 
and blue colors indicate the direction of current flow along clockwise and anti-clockwise directions corresponding 
to two opposite spins, respectively. The ring is penetrated by an Aharonov-Bohm (AB) flux $\Phi$ along its axis 
(upward direction). We describe our model by tight-binding (TB) Hamiltonian within non-interacting electron picture. 

The Hamiltonian for the entire system, ring with two side-attached leads, can be partitioned as,
\bea
\bm{H}=\left( \begin{array}{c c c}
\bm{H}_{\text{L}_1} & \bm{H}_{\text {L}_1\text{R}} & 0 \\
\bm{H}_{\text{L}_1\text{R}}^{\dag} & \bm{H}_{\text{R}} & \bm{H}_{\text{RL}_2} \\
0 & \bm{H}_{\text{RL}_2}^{\dag} & \bm{H}_{\text{L}_2}
\end{array}\right)
\eea
where, $\bm{H}_{\text {R}}$, $\bm{H}_{\text{L}_1}$ and $\bm{H}_{\text{L}_2}$ describe the Hamiltonian for the ring, 
lead-$1$ and lead-$2$, respectively. $\bm{H}_{\text{L}_{1(2)}\text{R}}$ represents the coupling between ring and lead-$1$($2$). 
%%%%%%%%%%%%%%%%%%%
%%%%%%%%%%%%%%%%%%%

The Hamiltonian for the $1$D ring can be written in Wannier basis as,
\beq
\bm{H}_{\text{R}}=\sum \limits_m  \bm{\epsilon}_m \mb{c}^{\dag}_m \mb{c}_m 
+\sum \limits_{m \ne n} \left( \mb{t}_{m,n} \mb{c}^{\dag}_{m} \mb{c}_{n} e^{i \Theta_{m,n}}+ \text{h.c.}\right)
\label{ham_ring}
\eeq
where,
\bea
&\mb{c}_{m}^{\dagger} =
\left ( \begin{array}{cc}
\mb{c}^{\dagger}_{m,\uparrow} & \mb{c}^{\dagger}_{m,\downarrow}
\end{array} \right ),\
\mb{c}_{m} =
\left ( \begin{array}{c}
\mb{c}_{m,\uparrow} \\ \mb{c}_{m,\downarrow}
\end{array} \right ).
\label{operators}
\eea
Here, $m$, $n$ are the site indices and $\uparrow$ ($\downarrow$) refers to spin index according to $S_z+$ and 
$S_z-$, respectively with the $z$-axis along the direction perpendicular to the plane of the ring. 
$\mb{c}^{\dag}_{m \sigma}$ ($\mb{c}_{m \sigma}$) is the creation (annihilation) operator at $m$-th site for an 
electron with spin $\sigma$. The on-site energy matrix is 
\beq
\bm{\epsilon}_m= \epsilon 
\left( \begin{array}{c c}
1 & 0 \\
0 & 1
\end{array}\right) 
\eeq
where $\epsilon$ is the on-site energy parameter set to zero. $\mb{t}_{m,n}$ is the long-range hopping integral 
between $m$-th and $n$-th sites. It has two parts corresponding to the bare hopping integral (\ie hopping due to 
bonding) and the hopping due to Rashba spin-orbit coupling. We write it as,
\beq
\mb{t}_{m,n}=\mb{t}^0_{m,n}+\mb{t}^{\prime}_{m,n}
\eeq
where the bare hopping part is given by,
\beq
\mb{t}^0_{m,n}=
t_{m,n}^l
\left( \begin{array}{c c}
1 & 0 \\
0 & -1 
\end{array}\right).
\label{tmn}
\eeq
In contrast to an ordinary hopping parameter inside the ring we consider an imaginary term for the long-range hopping 
integral expressed as~\cite{masuda2012interference},
\beq
t^l_{m,n}= \frac{i t(-1)^{m-n}} {(N/\pi) \sin{[\pi (m-n)/N]}}=t^{l~*}_{n,m}
\eeq 
with $t$, a real constant. Also in Eq.~(\ref{tmn}) down spin hopping integral is taken as negative with respect to that 
of the up spin hopping. This sign reversal for opposite spins along with the imaginary hopping is responsible for the 
helical behavior of the ring. It can also be interpreted as the time-reversal counter part as well. On the other hand, 
the contribution from the Rashba spin-orbit coupling is given by the hopping 
term~\cite{moca2005longitudinal,sheng2005spin,wang2008anisotropic},
\beq
\mb{t}^{\prime}_{m,n}=-i t_{rso} \left( \cos\phi_{m,n} \bm{\sigma}_x +\sin\phi_{m,n} \bm{\sigma}_y \right)
\eeq
with 
\beq
\phi_{m,n}=(\phi_m+\phi_n)/2,
\eeq
$\phi_m$ $=[2 \pi (m-1)/N$] being the azimuthal angle for $m$-th site. $t_{rso}$ is the strength of the hopping 
integral due to Rashba spin-orbit coupling. $\bm{\sigma}_{x(y)}$ is the Pauli spin matrix. The effect of magnetic 
flux is incorporated through the Peierl's phase factor $\Theta_{m,n}$ which can be expressed for the long-range 
hopping as~\cite{peierls1933theorie},
\beq
\Theta_{m,n}= \frac{2 \pi |m-n| \Phi}{N \Phi_0}
\eeq
$\Phi_0$ being the flux quantum.

Similar to the ring we write the Hamiltonian for the two $1$D leads in Wannier basis as,
\bea
\bm{H}_{\text{L}_{1(2)}}&=&\epsilon_0 \sum \limits_{m_{1(2)}} \mb{b}^{\dag}_{m_{1(2)}} \mb{b}_{m_{1(2)}} \non \\
&& +t_0 \sum \limits_{m_{1(2)}} \left(\mb{b}^{\dag}_{m_{1(2)}} \mb{b}_{m_{1(2)}+1}+ \text{h.c.}\right)
\eea
where $\epsilon_0$ and $t_0$ are the on-site energy and nearest-neighbor hopping integral for the leads. We have 
used another notation for the creation ($\mb{b}^{\dag}_{m_{1(2)}}$) and annihilation operators ($\mb{b}_{m_{1(2)}}$) 
for the leads. They are expressed in similar way as that of $\mb{c}^{\dag}_{m_{1(2)}}$ and $\mb{c}_{m_{1(2)}}$ (see 
Eq.~(\ref{operators})). The ring-to-lead couplings for the two leads are described as,
\beq
\bm{H}_{\text{L}_{1(2)}{\text{R}}}= \tau \left( \bm{c}^{\dag}_{m} \bm{b}_{m_{1(2)}}+ \text{h.c.} \right).
\eeq
$m$ and $m_{1(2)}$ are indices corresponding to the neighboring sites situated at the boundaries of the ring and 
lead-$1$($2$). $\tau$ is the coupling strength between the lead and the ring.

%---------------------------------------------------------------------------------
\section{Calculation of Transmission probability: Green's function formalism}
\label{sec3}
%---------------------------------------------------------------------------------

In order to calculate the transmission probability of the incoming electron through the helical ring we adopt 
Green's function formalism~\cite{datta1997electronic} which is summarized below.

First, let us define the single-particle retarded (advanced) Green's function for our model (helical ring with 
two side-attached leads) as,
\beq
\bm{G}^{r(a)}=\left(z^{\pm} \bm{I} -\bm{H}\right)^{-1} 
\eeq
where $z^{\pm}=(E\pm i\eta)$ and $\eta\rightarrow 0^+$. $E$ is the incoming electron energy. Similar to the 
Hamiltonian, we can also partition the Green's function for the entire system corresponding to the different 
parts of it like, 
\bea
\bm{G}^r=\left( \begin{array}{c c c}
\bm{G}_{\text{L}_1} & \bm{G}_{\text {L}_1\text{R}} & 0 \\
\bm{G}_{\text {L}_1{\text{R}}}^{\dag} & \bm{G}_{\text{R}} & \bm{G}_{\text{RL}_2} \\
0 & \bm{G}_{\text{RL}_2}^{\dag} & \bm{G}_{\text{L}_2}
\end{array}\right).
\eea
Now, we map $\bm{G}^r$ corresponding to the full Hilbert space of the entire system to the reduced Hilbert space 
which consists of the ring alone. This allows us avoiding the calculation of the infinite-dimensional Green's 
function for the full system comprised of finite dimensional ring and two semi-infinite leads. After calculation 
we have the effective Green's function for the isolated ring as follows,
\beq
\bm{\mathcal{G}}^r=\left(z^+\bm{I}-\bm{H}_R-\bm{\Sigma}_{\text{L}_1}^r-\bm{\Sigma}_{\text{L}_2}^r\right)^{-1}.
\label{effectiveG}
\eeq
where
\beq
\bm{\Sigma}_{\text{L}_{1(2)}}^r=\bm{H}_{\text{L}_{1(2)}\text{R}}^{\dag} \bm{g}_{\text{L}_{1(2)}}^r \bm{H}_{\text{L}_{1(2)}R}.
\eeq
$\bm{g}^r_{\text{L}_{1(2)}}$ is the Green's function for the lead-$1$($2$) defined as,
\beq
\bm{g}_{\text{L}_{1(2)}}^r=\left(z^+\bm{I}-\bm{H}_{\text{L}_{1(2)}}\right)^{-1}.
\eeq
Thus, dimension of the effective Green's function is same as that of the ring alone. Now, it is sufficient to consider 
only the isolated ring with two modified boundary sites characterized by effective site-potentials. After the 
mapping, we have two terms $\bm{\Sigma}_{\text{L}_1}^r$ and $\bm{\Sigma}_{\text{L}_2}^r$. They are the retarded 
self-energies responsible for the lead-$1$-to-ring and ring-to-lead-$2$ couplings, respectively. They have 
non-zero elements only for the boundary sites where the lead(s) and the ring are connected to each other. The 
self-energy for each lead is expressed in terms of the bare Green's function of the corresponding lead. We 
calculate them considering both the leads as single channel semi-infinite periodic chains having only nearest 
neighbor hopping. Expressions for those self-energies in terms of the incoming electron energy and the hopping
strengths are given by~\cite{datta1997electronic},
\bea
&& \bm{\Sigma}_{L_{\alpha}~m\sigma,m^{\prime}\sigma^{\prime}}^r (E) \non \\ 
&= &\frac{\tau^2}{2t_0^2} \delta_{mm_{\alpha}} 
\delta_{m^{\prime}m_{\alpha}} \delta_{\sigma \sigma^{\prime}} 
\left[z^+-\epsilon_0-i\sqrt{4 t_0^2-(z^+-\epsilon_0)^2}\right].\non \\
\eea
Here, $\alpha$ is the lead index. $m$ and $m_{\alpha}$ are the site indices in the ring and $\alpha$-th lead, 
respectively. Note that, the self-energies corresponding to the two leads have both finite real and imaginary 
parts contributing to the effective Green's function. The real part is responsible for the energy level shifting 
whereas, the imaginary part describes the broadening of the levels. In other words, one can separate out the 
broadening matrix parts corresponding to each $\sigma$ like,
\beq
\bm{\Gamma}_{\text{L}_{1(2)}}^{\sigma}= -2 \text{Im} \left(\bm{\Sigma}^{r~\sigma}_{\text{L}_{1(2)}}\right).
\eeq 

Now, in terms of the broadening matrices and the effective Green's functions, the transmission function can be found
from the following relation~\cite{datta1997electronic},
\beq
T_{\sigma \sigma^{\prime}} (E) =\text{Tr} \left[ 
\bm{\Gamma}_{\text{L}_1}^{\sigma}.\bm{\mathcal{G}}^r.\bm{\Gamma}_{\text{L}_2}^{\sigma^{\prime}}.\bm{\mathcal{G}}^a \right].
\label{trans}
\eeq
$T_{\sigma \sigma^{\prime}}$ represents the transmission probability of the incoming electron through lead-$1$ with 
spin $\sigma$ as an electron with spin $\sigma^{\prime}$ into lead-$2$. For the advanced part of the effective 
Green's function, one can take the complex conjugate of the expression written in Eq.~(\ref{effectiveG}).
%-------------------------------
\section{Numerical Results}
\label{nr}
%-------------------------------

For numerical calculation, we use the unit where $c=h=e=1$. The hopping parameter $t$ is taken as the 
unit of energy. Throughout our calculation we take the ring parameter values as, $N=91$, $t=2$, $a=1$ and for clean 
system $\epsilon=0$; the lead parameters as, $\epsilon_0=0$, $t_0=3$. The leads are attached to the sites $n_{\text{L}_1}$ 
and $n_{\text{L}_2}$, respectively with the coupling strength $\tau=2.5$ for both leads signifying strong 
lead-to-ring coupling. We set them as, $n_{\text{L}_1}=1$ and $n_{\text{L}_2}=46$. Now, we illustrate our results in three 
different sub-sections in order to describe the effect of magnetic flux, Rashba SOI and scalar disorder successively.
%--------------------------------------------------
\subsection{Effect of magnetic flux}
%--------------------------------------------------

\begin{figure}[!thpb]
\centering
{\includegraphics[width=0.98\linewidth]{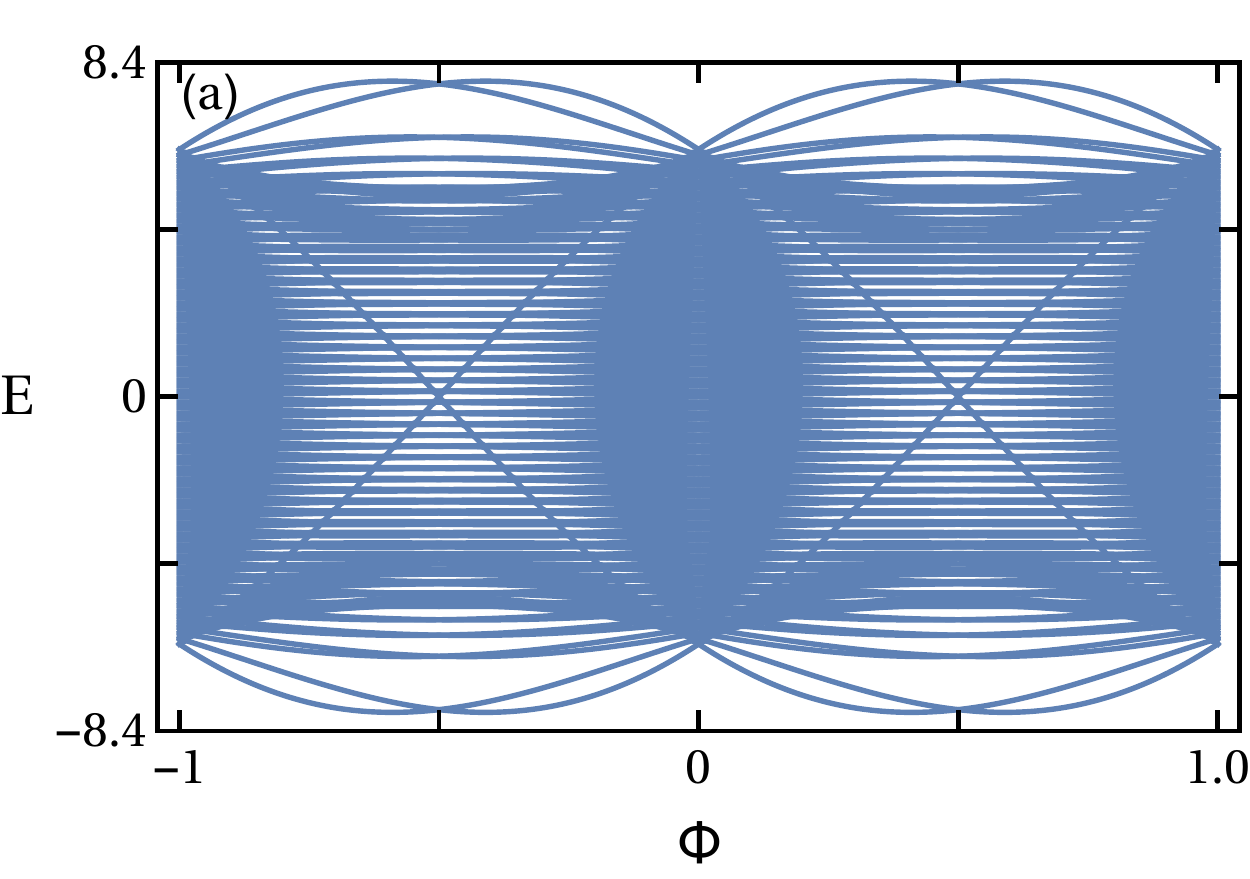}}\\
{\includegraphics[width=0.492\linewidth]{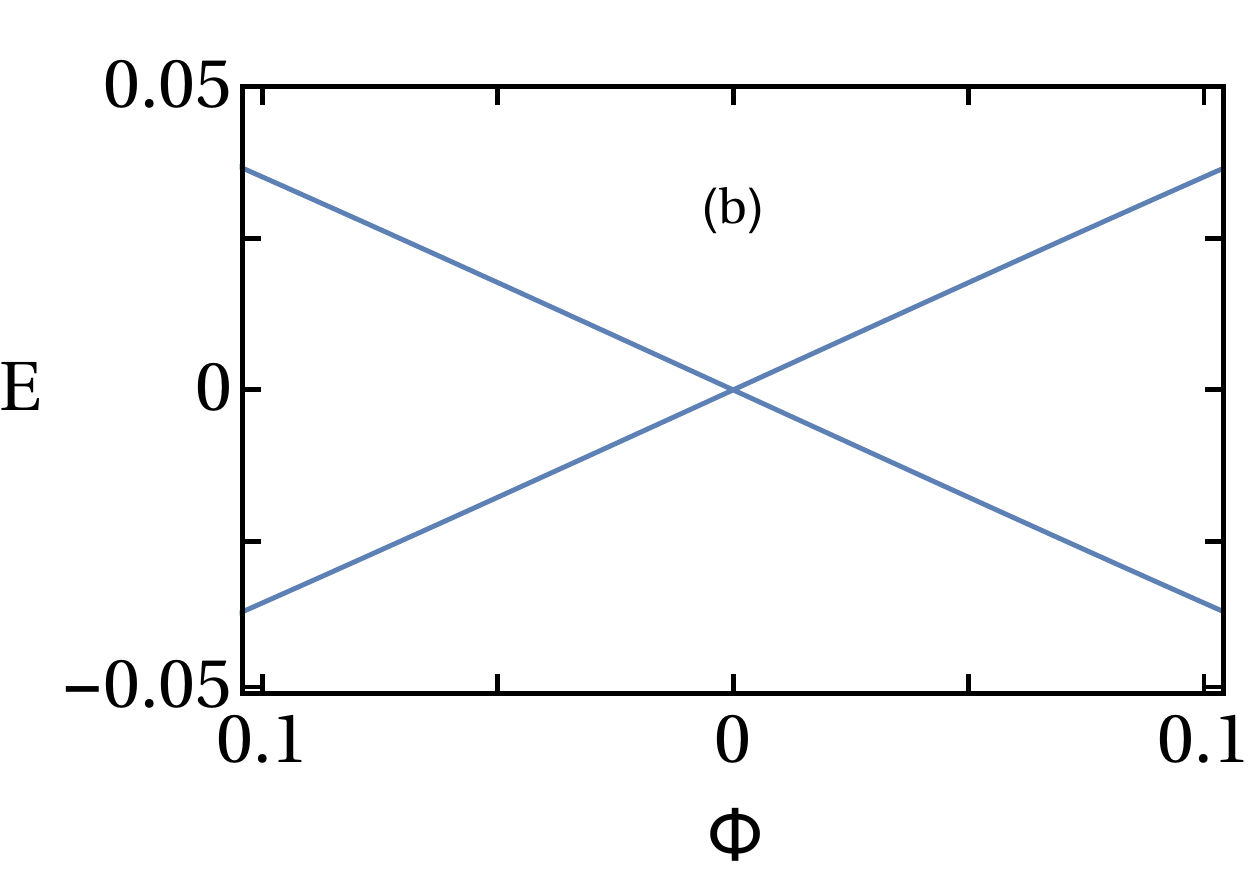}
\includegraphics[width=0.492\linewidth]{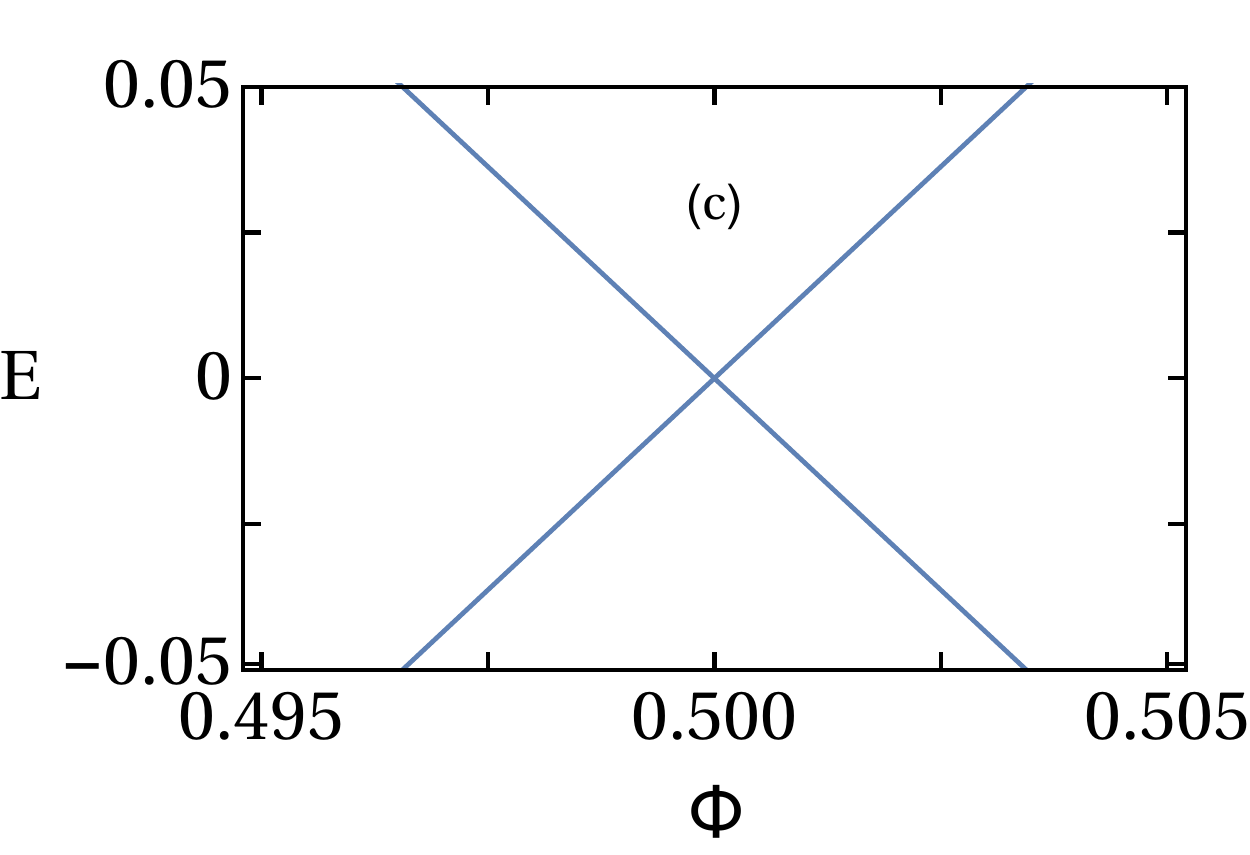}}
\caption{(Color online) Upper panel: (a) Full energy spectrum ($E$-$\Phi$) of a clean ring with atomic sites $N=91$. 
Lower Panel: The zoomed-in versions of the full spectrum around (b) $\Phi=0$ and (c) $\Phi=\Phi_0/2$.}
\label{spec}
\end{figure}

In Fig.~\ref{spec} we plot the energy levels of the ring as a function of AB flux $\Phi$. We take an isolated ring 
having $91$ number of lattice sites and diagonalize the Hamiltonian to find the energy levels. Fig.~\ref{spec}(a) 
represent the entire spectrum of the isolated ring. We zoom-in the spectrum around  $\Phi=0$ and $\Phi=\Phi_0/2$ and 
present them in Fig.~\ref{spec}(b) and Fig.~\ref{spec}(c), respectively. 

The most remarkable feature is that there is zero energy crossing at $\Phi=0$. Similar zero energy crossing in the 
energy spectrum of this helical ring model was also articulated 
%%%%%%%%%%%%%%%%%%%%%%%%%%%%%%%%
%%%%%%%%%%%%%%%%%%%%%%%%%%%%%%%%
in Ref.~\onlinecite{masuda2012interference} in absence of magnetic flux. This zero energy crossing associated with 
two opposite spin states is analogous to the linear dispersion relation of the topological insulator 
edge states~\cite{kane2005z,bernevig2006quantum,konig2007quantum}. 
An analytic derivation of the dispersion relation ($E$ vs. $k$) of the isolated ring in presence of AB flux is 
presented in Appendix. 

In addition to $\Phi=0$ we get similar zero energy crossings at all integer multiples of half-flux quantum values of 
the applied AB flux. Due to the crossing we have both the positive and negative slopes of the curves \ie both positive 
%%%%%%%%%%%%%%%%%%%%%%%%%%%%%%%%
\begin{figure*}[!thpb]
\centering
\includegraphics[height=09.7cm,width=0.92\linewidth]{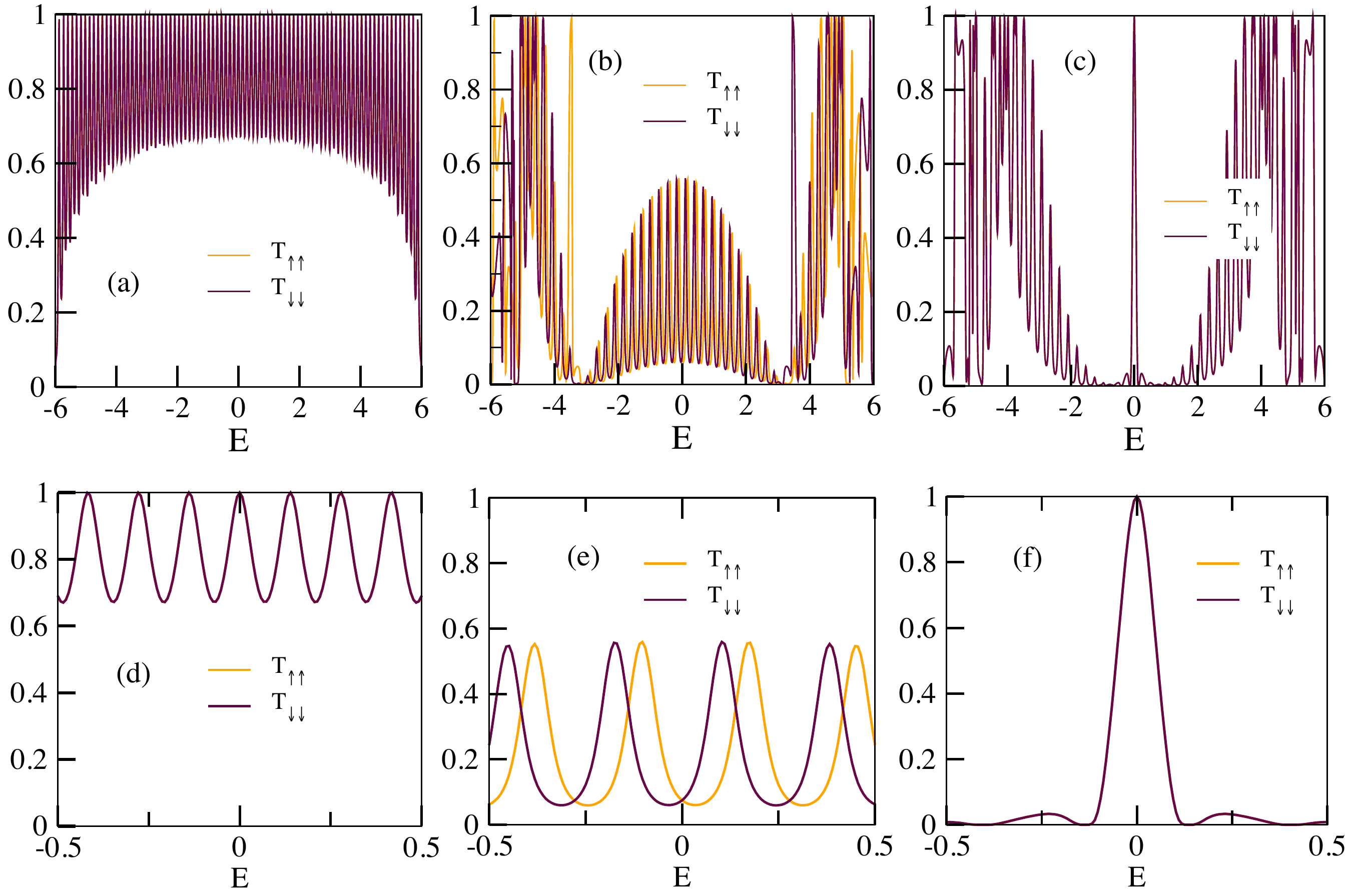}
\caption{(Color online) Quantum-mechanical transmission probabilities $T_{\uparrow\uparrow}$ (orange color) and 
$T_{\downarrow\downarrow}$ (maroon color) are plotted as a function of incoming electron energy $E$ corresponding 
to a ring with $N=91$ for three values of AB flux (a) $\Phi=0$, (b) $\Phi=\Phi_0/4$ and (c) $\Phi=\Phi_0/2$. (d), 
(e) and (f) represent the zoomed-in versions of (a), (b) and (c) respectively around $E=0$.} 
\label{te_phi}
\end{figure*}
%%%%%%%%%%%%%%%%%%%%%%%%%%%%%%%%
and negative group velocities are possible for the electrons with two opposite spins and this results in appearance 
of counter-propagating currents characterized by two opposite spin states. To ensure the current direction 
corresponding to two spins one can calculate the bond current between two sites along any arm of the ring considering 
the up and down spin electrons~\cite{dutta2012magnetic}. We skip this part as one can easily guess the two opposite 
signs of the currents by looking at the signs of the hopping integral as given in Eq.~(\ref{tmn}). 

To investigate the outcome of the helical states with the applied magnetic flux we study the transmission probability 
of electrons through the ring. We compute the transmission probability of electrons following Eq.~(\ref{trans}) for both
up and down spin electrons and show their behaviors in Fig.~\ref{te_phi} with respect to the incoming electron energy 
$E$. The maroon and orange color represent $T_{\uparrow\uparrow}$ and $T_{\downarrow\downarrow}$ \ie the probability 
of transmission of the $\uparrow$ and $\downarrow$ spin electrons without any spin-flip scattering. Spin flipping is not 
possible as we have not included the Rashba spin-orbit interaction so far. Here, Figs.~\ref{te_phi}[(a), (b) and (c)] 
correspond to the three different values of the AB flux $\Phi=0$, $\Phi_0/4$ and $\Phi_0/2$, respectively. We zoom into 
all the three figures ((a)-(c)) around $E=0$ and display them in Figs.~\ref{te_phi} [(d), (e) and (f)], respectively.

In absence of any flux ($\Phi=0$) we note that transmission curve shows oscillatory behavior (see Fig.~\ref{te_phi}(a)). 
The oscillation is much more prominent in the zoomed-in version (see Fig.~\ref{te_phi}(d)). Our model ring is attached 
to two leads. So, the wave functions corresponding to the incoming electrons pass through the two arms of the ring. 
After traveling through the two arms they again meet at the junction, where the lead-$2$ is attached to the ring, either 
constructively or destructively. This quantum interference leads to an oscillatory behavior of the transmission 
probability. In the oscillation, number of peaks describing the resonances is equal to the number of energy levels of 
the isolated ring. To be mentioned, we consider completely unpolarized beam for the incoming electrons.  

However, this phenomenon of quantum interference is true for both the up and down spin electrons. Only, the difference 
between them is that they travel along two opposite directions within the ring. The transmission probabilities 
corresponding to two spin states of the electrons are exactly superposed on each other as the time-reversal 
symmetry between the $\uparrow$ and $\downarrow$ spin states is protected. Our result for $\Phi=0$ (see Fig.~\ref{te_phi}(d)) 
corroborates with the results obtained in Ref.~\onlinecite{masuda2012interference} where the energy dependence of 
transmission probability of electrons is reported. Although the frequency of the oscillation being different as it depends on 
the ring size. However, the most interesting feature of the transmission spectra is that there are zero-energy peaks 
for both the spin states and these peaks correspond to the zero energy crossing in the energy spectra at $\Phi=0$ as 
shown in Fig.~\ref{spec}(b).

Now, as soon as we introduce AB flux $\Phi$, the time reversal symmetry between the up and down spin states breaks 
down \ie the symmetry between left-moving up spin and right-moving down spin electrons no longer exists. A phase 
difference is introduced between the wave functions corresponding to the clockwise and anti-clockwise propagating 
electrons with opposite spins. As a result, in the feature of transmission probabilities the oscillations corresponding 
to $T_{\uparrow\uparrow}$ and $T_{\downarrow\downarrow}$ are different from each other by a phase factor (see 
Fig.~\ref{te_phi}(b)). However, the amplitudes of oscillation are not same throughout the energy window because of 
the destructive interference between the wave functions propagating along the two arc-lengths. The removal of the 
degeneracy between $\uparrow$ and $\downarrow$ is much more prominent in Fig.~\ref{te_phi}(e). This leads to a 
possibility of obtaining finite spin polarization depending on the energy value and applied AB flux. It should be noted, there 
is no zero-energy peak in the transmission spectra for this flux value. This can be understood by looking at the energy 
spectra around this flux value as depicted in Fig.~\ref{spec}(a).
 
Nonetheless, this phenomenon of separation of the transmission curves corresponding to the two different spin states 
by a phase difference is true as long as the flux value is different from the half-flux quantum. To investigate this, 
in Fig.~\ref{te_phi}(c) we plot the probability of $\uparrow$ and $\downarrow$ spin electron transmission vs. energy 
$E$ for $\Phi=\Phi_0/2$ choosing all the other parameter values same as in the previous two cases for other two 
flux ($\Phi$) values. We observe that when $\Phi=\Phi_0/2$, $T_{\uparrow\uparrow}$  and $T_{\downarrow\downarrow}$ are 
again exactly superposed on each other nullifying the possibility of getting finite spin polarization. 
Moreover, we have a peak at exactly $E=0$ and this peak corresponds to the crossing at zero energy for $\Phi=\Phi_0/2$ 
as shown in Fig.~\ref{spec}(c). The persistence of the zero energy peak correspond to the spinor nature of the electrons. 
However, the appearance of the zero-energy peaks at particularly these flux values (\ie $\pm \Phi_0/2$) 
is completely a topological signature of the helical ring with the underlying inversion symmetry. They neither depend on the 
details (different sizes of the two ring arms) of the ring nor the position of the attached leads. Note that, there is no 
Zeeman effect in our case. We only consider a single channel of the ring. Zeeman splitting does not play any role as far as the 
radius of the ring is less than the radius of the cyclotron orbit. In our case, only the AB effect is the effective one.

\begin{figure}[!thpb]
\centering
\includegraphics[height=5.2cm,width=1.00\linewidth]{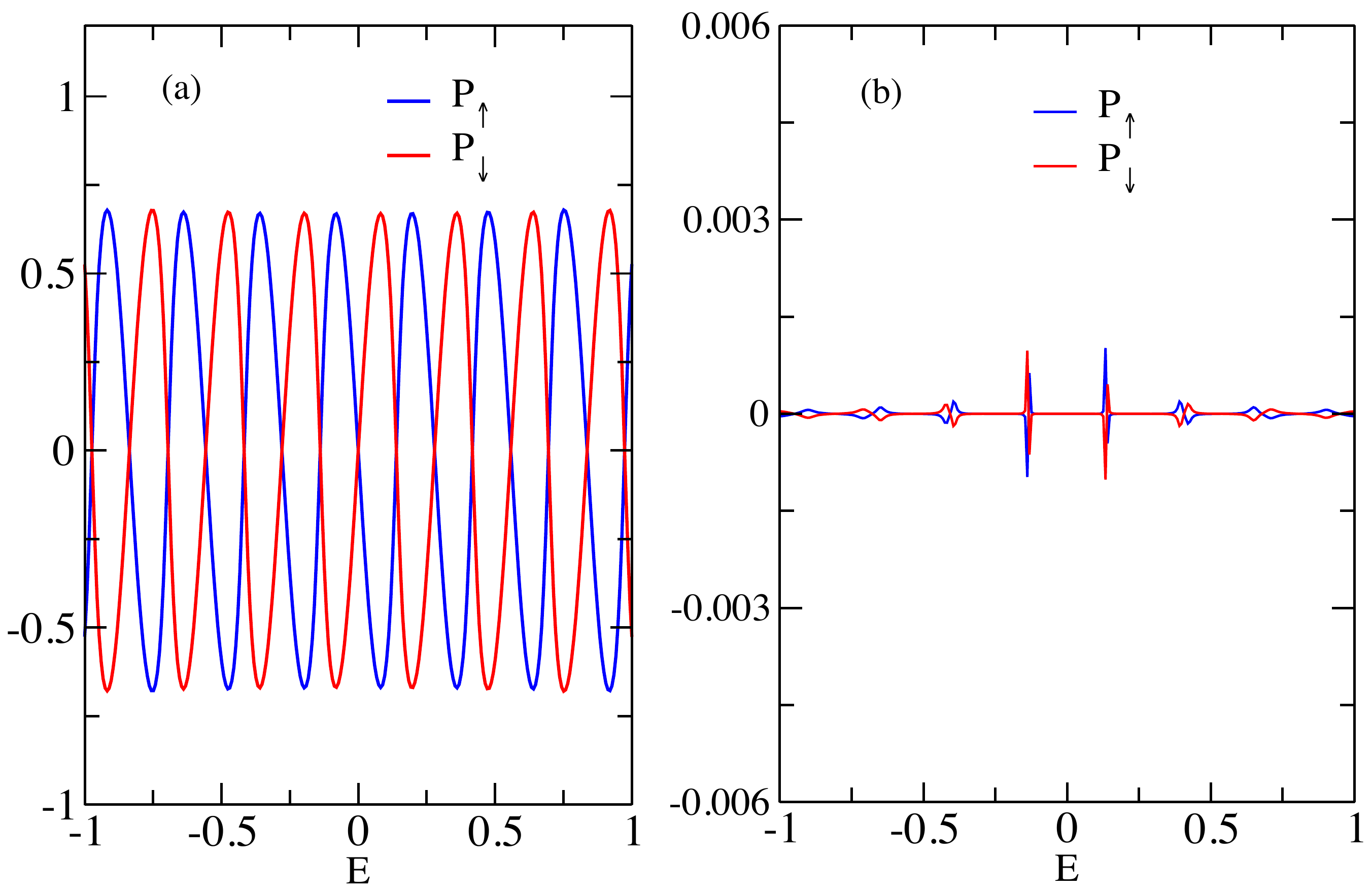}
\caption{(Color online) Spin polarization is shown as a function incident electron energy $E$ for up ($P_{\uparrow}$) 
and down ($P_{\downarrow}$) spin electrons with AB flux (a) $\Phi=\Phi_0/4$ and (b) $\Phi_0/2$ respectively.}
\label{pol}
\end{figure}

To analyze the spin polarization in detail we define a function, namely the spin polarization factor
as~\cite{moldoveanu2010tunable},
\beq
P_{\uparrow(\downarrow)}=\frac{T_{\uparrow\uparrow(\downarrow\downarrow)}-T_{\downarrow\downarrow(\uparrow\uparrow)}}
{T_{\uparrow\uparrow}+T_{\downarrow\downarrow}}
\eeq
where $P_{\uparrow}$ and $P_{\downarrow}$ correspond to the two different spin states $\uparrow$ and $\downarrow$. Their 
behaviors with respect to the incoming electron energy are shown in Fig.~\ref{pol} where panel (a) and (b) correspond to 
$\Phi=\Phi_0/4$ and $\Phi_0/2$ respectively. The blue and red lines represent the efficiency factors for $\uparrow$ and 
$\downarrow$ spin electrons, respectively. From Fig.~\ref{pol}(a) we see that both $P_{\uparrow}$ and $P_{\downarrow}$ have 
oscillatory behaviors being dual to each other. Their amplitudes 
%%%%%%%%%%%%%%%%%%%%%%%%%%%%%%
%%%%%%%%%%%%%%%%%%%%%%%%%%%%%%%%
run between $0.7$ and $-0.7$ approximately whereas the maximum allowed values for the amplitudes are $\pm 1$. Looking 
at the expressions for the efficiency factors we can say that $P_{\uparrow}=1$ indicates that $T_{\downarrow\downarrow}$ 
is exactly equal to zero but $T_{\uparrow\uparrow}=1$. That means we have only $\uparrow$ spin electron transmission. 
The transmissions of $\downarrow$ spin electrons are completely blocked for that particular energy value. In contrast, 
when $P_{\downarrow}=1$ we have exactly the opposite scenario. However, in our case amplitude of the $P_{\sigma}$ curve 
is almost $0.7$ \ie all the energy values where $P_{\downarrow}=0.7$ we have $70\%$ polarization with 
$\downarrow$ spin being favorable one, suppressing $\uparrow$ spin transmission and vice-versa. For the rest of the energy 
values the spin polarization is less than $70\%$.

From the oscillations of $P_{\sigma}$ ($\sigma$ may be $\uparrow$ or $\downarrow$) we obtain some energy values where 
transition happens \ie the efficiency factors change their signs. Now, the change in sign of one efficiency factor, 
say $P_{\sigma}$, signifies that the transmission of electrons having opposite spin starts dominating. Note that, we have 
shown the result for a particular energy range. Although for the entire energy band we get similar features of 
$P_{\uparrow}$ and $P_{\downarrow}$. Therefore, for a particular flux value the spin polarization is different for 
different energy values or in other words, spin polarization of the helical ring is energy-dependent. 

On the other hand, if we tune the flux we get different behaviors of the efficiency factors for a particular energy value. 
To examine this, we calculate the efficiency factors for $\Phi=\Phi_0/2$. The up and down spin electron transmission again 
become exactly symmetric to each other resulting in vanishing spin polarization at this flux value. This is 
clear from Fig.~\ref{pol}(b) where both $P_{\uparrow}$ and $P_{\downarrow}$ are vanishingly small ($\sim10^{-4}$) in magnitude. 
However, the deviation of the efficiency value ($P_{\uparrow}$ or $P_{\downarrow}$) from exact zero occurs 
due to the finite size of the ring.

The most striking feature is that we obtain finite spin polarization using the long-range hopping model just by tuning 
the magnetic flux. If we fix the magnetic flux to a finite value other than half-flux quantum we get finite spin polarization
by tuning the applied bias. In other words, we can tune the magnetic flux to have the spin polarization finite or zero for a 
particular energy value. This is in contrast to the behavior of ordinary periodic ring. In ordinary ring with nearest-neighbor 
hopping, when we apply magnetic flux the transmission is modified as a result of the quantum interference of the electronic 
wave functions traveling along the two arms of the ring. However, one cannot break the spin degeneracy just by tuning the 
applied AB flux. So, the separation between the up and down spin transmission is not possible~\cite{moldoveanu2010tunable}. 
They are exactly symmetric to each other.

%--------------------------------------------------
\subsection{Effect of Rashba spin-orbit interaction}
\begin{figure}[!thpb]
\centering
\includegraphics[height=4.7cm,width=0.575\linewidth]{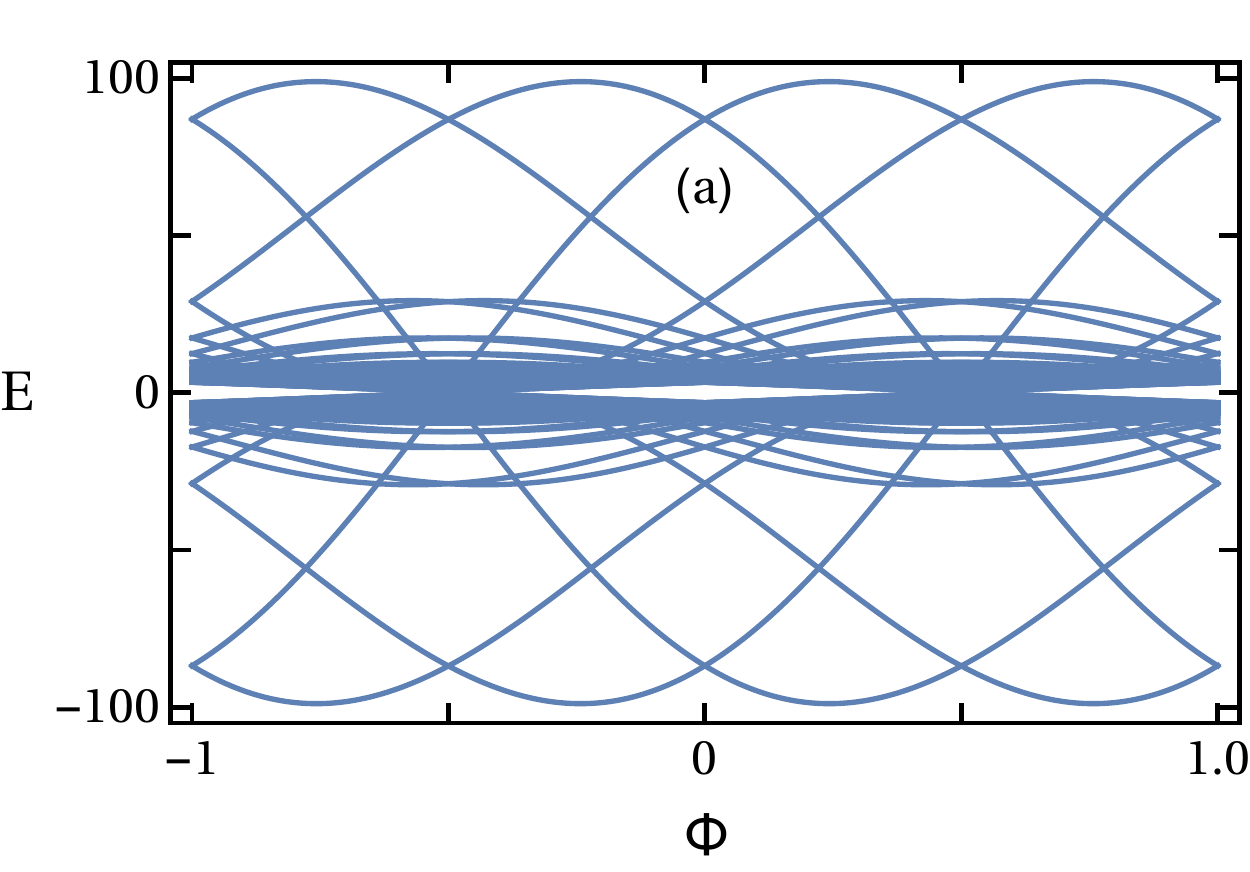}
\includegraphics[height=4.8cm,width=0.407\linewidth]{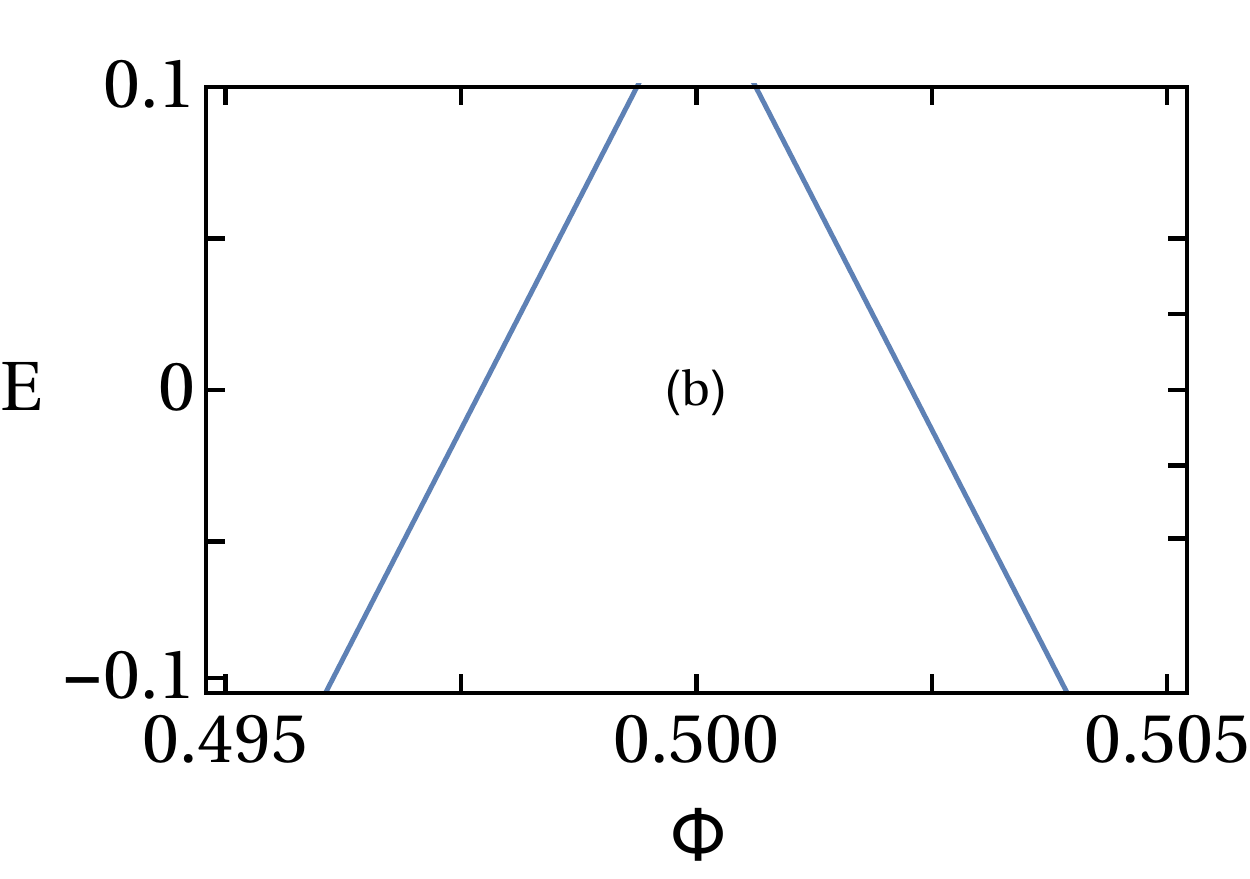}
\caption{(Color online) Left column (a): Full energy spectrum ($E$-$\Phi$) of a clean ring ($N=91$) in presence of RSOI 
($t_{rso}=1.5$). Right column (b): Part of the spectrum of Fig.~\ref{spec_rashba}(a) while zoomed-in around $\Phi=\Phi_0/2$.}
\label{spec_rashba}
\end{figure}

%--------------------------------------------------
In this sub-section we investigate the effect of RSOI on the transport properties of the helical ring with long-range 
hopping. Fig.~\ref{spec_rashba} displays the energy spectrum ($E$-$\Phi$) of the isolated ring with $N=91$ in presence 
of RSOI with strength $t_{rso}=1.5$. We observe that there is a gap at the central region (\ie around $\Phi=0$) of the 
spectrum. In absence of RSOI we have a zero-energy crossing at $\Phi=0$ and the crossing no longer exists in presence 
of Rashba spin-orbit interaction. This gap arises due to spin flip scattering which makes this helical ring to behave like an 
%%%%%%%%%%%%%%%%%%%%%%%%%%%%%%%%
%%%%%%%%%%%%%%%%%%%%%%%%%%%%%%%%
ordinary insulator at $\Phi=0$. Such gap may seem to appear due to the destructive interference caused by 
the incomplete Rashba spin flipping along the chain. Also, constructive intereference is possible for very small size of the
ring  and this can lead to the closing of the gap around $\Phi=0$. To overcome this effect we have taken more than $75$ 
number of sites in the ring.
We obtain wider energy band and this enhancement can be realized more prominently by comparing 
with Fig.~\ref{spec}. The splitting for outer bands is higher compared to that of the inner bands. We also zoom-in 
the spectrum around $\Phi=\Phi_0/2$. We have energy levels at $E=0$ corresponding to which we expect finite 
transmission but there is no crossing at this flux value similar to the case of $\Phi=0$. In topological insulator 
or the spin quantum Hall states we find helical edge states that disperse linearly. In these systems Rashba 
spin-orbit interaction is an inherent property. In our $1$D ring with long-range hopping we see that such Dirac-like 
crossings do not exist in spectrum in presence of RSOI. When we incorporate Rashba spin-orbit interaction all the 
energy levels get split. This results in a gap in the spectrum. For finite values of $\Phi$, we have two 
contributions to the phase part, one due to the magnetic flux and another due to Rashba spin-orbit interaction. 
There may be compensation of the phase due to one by the other resulting in a shifting of energy levels. Therefore, 
we loss the zero energy crossing at the half flux quantum values also. 

To be noted, a minute asymmetry ($\sim 0.05\%$) appears around $E=0$ in Fig.~\ref{spec_rashba}(b). This asymmetry is 
fully due to the finite-size effect of the ring which causes an overall shifting of the energy spectra from the zero 
energy value. It does not depend on the flux value. The particle-hole symmetry is preserved in our system even in 
presence of RSOI.

\begin{figure}[!thpb]
\centering
\includegraphics[width=0.999\linewidth]{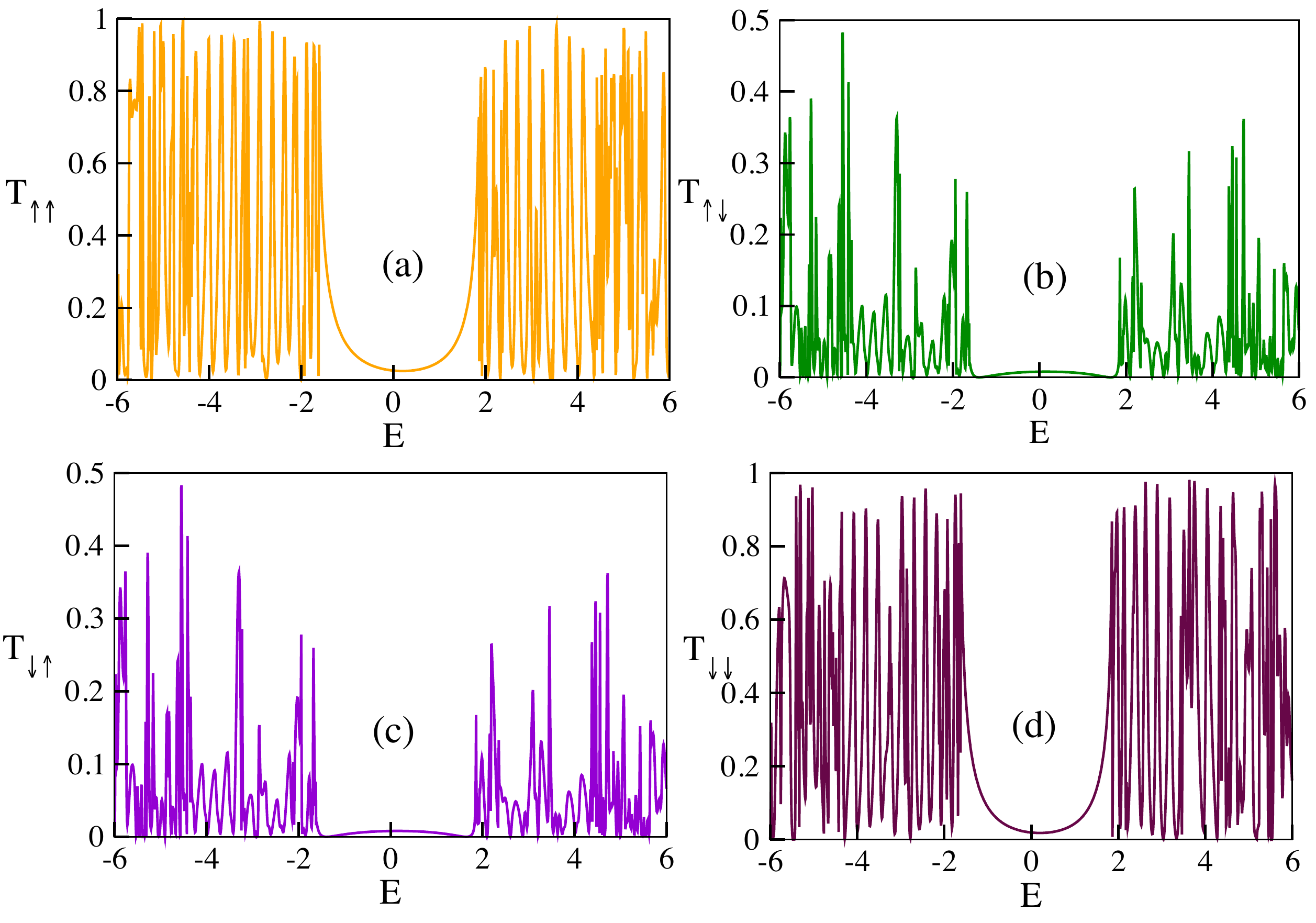}
\caption{(Color online) Plot of transmission probabilities ((a) $T_{\uparrow\uparrow}$, (b) $T_{\uparrow\downarrow}$, 
(c) $T_{\downarrow\uparrow}$, and (d) $T_{\downarrow\downarrow}$) as a function of incoming electron energy $E$ for a 
ring with $N=91$ in presence of RSOI ($t_{rso}=1.5$) and  $\Phi=\Phi_0/4$.}
\label{te_Rphi25}
\end{figure}

Now, we study the transmission phenomena in presence of RSOI. In Fig.~\ref{te_Rphi25} we show all four possible 
%%%%%%%%%%%%%%%%%%%%%%%%%%%%%%%%
%%%%%%%%%%%%%%%%%%%%%%%%%%%%%%%%
spin transmission probabilities as a function of injecting electron energy $E$ in presence of magnetic flux 
$\Phi=\Phi_0/4$. Fig.~\ref{te_Rphi25}[(a), (b), (c) and (d)] correspond to $T_{\uparrow\uparrow}$ (orange lines), 
$T_{\uparrow\downarrow}$ (green lines),  $T_{\downarrow\uparrow}$ (violet lines) and $T_{\downarrow\downarrow}$ (maroon lines) respectively.
Now, we see that the probability of $\uparrow$ and $\downarrow$ spin electron transmissions without any spin flipping 
are comparable in magnitude but they are not exactly equal to each other. There are central gaps around $E=0$ in the 
both spectra of $T_{\uparrow\uparrow}$ and $T_{\downarrow\downarrow}$. With the increase of Rashba hopping strength
these gap-widths increase. On the other hand, due to presence of RSOI we have finite spin-flip transmission 
probability for both up and down spin electrons as shown in Fig.~\ref{te_Rphi25}(b) and Fig.~\ref{te_Rphi25}(c) respectively.
Note that, the magnitude of spin-flip transmission probability is much smaller compared to the spin-conserving transmission 
probabilities. Also, there are very small probability of transmission of electrons at $E=0$ for this flux value.

RSOI actually behaves like an effective magnetic field which depends on the momenta of the electrons. It causes 
a phase difference between the electronic wave functions corresponding to the two spin states traveling along 
opposite directions. This induced phase difference modifies the quantum interference phenomenon both constructively 
as well as destructively. As a consequence, we get different magnitudes of the transmission probabilities for any 
particular energy value. It is evident while we compare Figs.~\ref{te_phi}[(a)-(c)] and Figs.~\ref{te_Rphi25}.
%%%%%%%%%%%%%%%%%%%%%%%%%%%%%%%%
\begin{figure}[!thpb]
\centering
\includegraphics[width=1.0\linewidth]{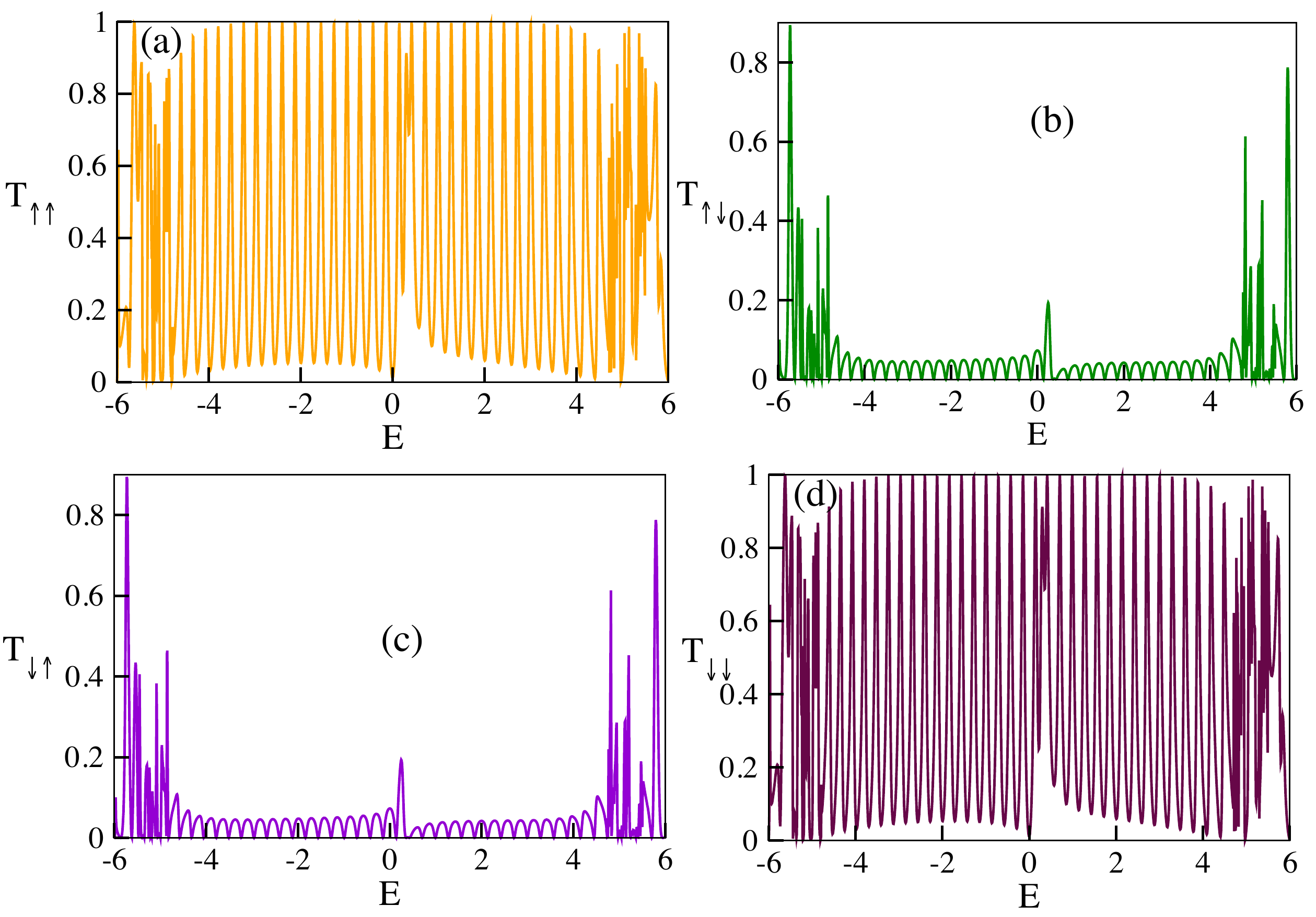}
\caption{(Color online) Quantum-mechanical transmission probabilities (a) $T_{\uparrow\uparrow}$, 
(b) $T_{\uparrow\uparrow}$, (c) $T_{\uparrow\uparrow}$, and (d) $T_{\downarrow\downarrow}$ are shown as a function 
of incoming electron energy $E$ for $\Phi=\Phi_0/2$. The values of other parameters of the ring are kept same as in 
Fig.~\ref{te_Rphi25}.}
\label{te_Rphi5}
\end{figure}
%%%%%%%%%%%%%%%%%%%%%%%%%%%%%%%%
Whereas, for $\Phi=\Phi_0/2$, we have already seen from the energy spectra that there are energy levels available at 
$E=0$. This reflects in the transmission spectra accordingly. We have finite spin-flip transmission at this energy 
value. In Fig.~\ref{te_Rphi5} we display the four transmission probabilities of electrons through the ring keeping 
all the parameter values same as in Fig.~\ref{te_Rphi25}. It looks that the central gaps of the spectra of 
$T_{\uparrow\uparrow}$ and $T_{\downarrow\downarrow}$ disappear for $\Phi=\Phi_0/2$. If we observe carefully then 
we notice that there are dips at $E=0$ in the spectra of electron transmission without spin-flipping. If we change 
the RSOI strength there may be finite spin transmission at this energy value in place of a central gap. We can say 
that the zero energy peak no longer exist when $\Phi=\Phi_0/2$ also. Additionally, $T_{\uparrow\uparrow}$ and 
$T_{\downarrow\downarrow}$ become exactly symmetric to each other. On the other hand, spin-flip electron transmission 
is finite throughout the energy window. Therefore, in presence of RSOI zero-peaks associated with the zero energy 
crossings are absent for any value of magnetic flux.

From all the figures of the transmission probabilities in presence of RSOI we can argue that the possibility
of getting finite spin polarization is extremely small. We do not plot the spin polarization here. 
On the contrary, in case of ordinary ring one can design a spin-filter device based on the spin polarization just by 
tuning the RSOI alone~\cite{moldoveanu2010tunable}. Note that, the small asymmetry around $E=0$ in each plot of the 
transmission spectra appears because of the asymmetry in the energy spectra originating due to the finite size of the 
discrete model.

%--------------------------------------------------
\subsection{Effect of disorder}
%--------------------------------------------------
The helical states obtained at the edges of a $2$D topological insulator are robust to non-magnetic 
impurity~\cite{qi2010quantum,hasan2010colloquium,sczhangreview}. Here we investigate whether the electron flows 
corresponding to the opposite spins traveling in opposite directions along each arm of the ring get affected by static 
disorder or not. For this, we introduce static scalar random disorder into the system. This type of 
disorder does not affect the spin state of electrons. We 
%%%%%%%%%%%%%%%%%%%%%%%%%%%%%%%%
\begin{figure}[!thpb]
\centering
\includegraphics[height=4.5cm,width=0.57\linewidth]{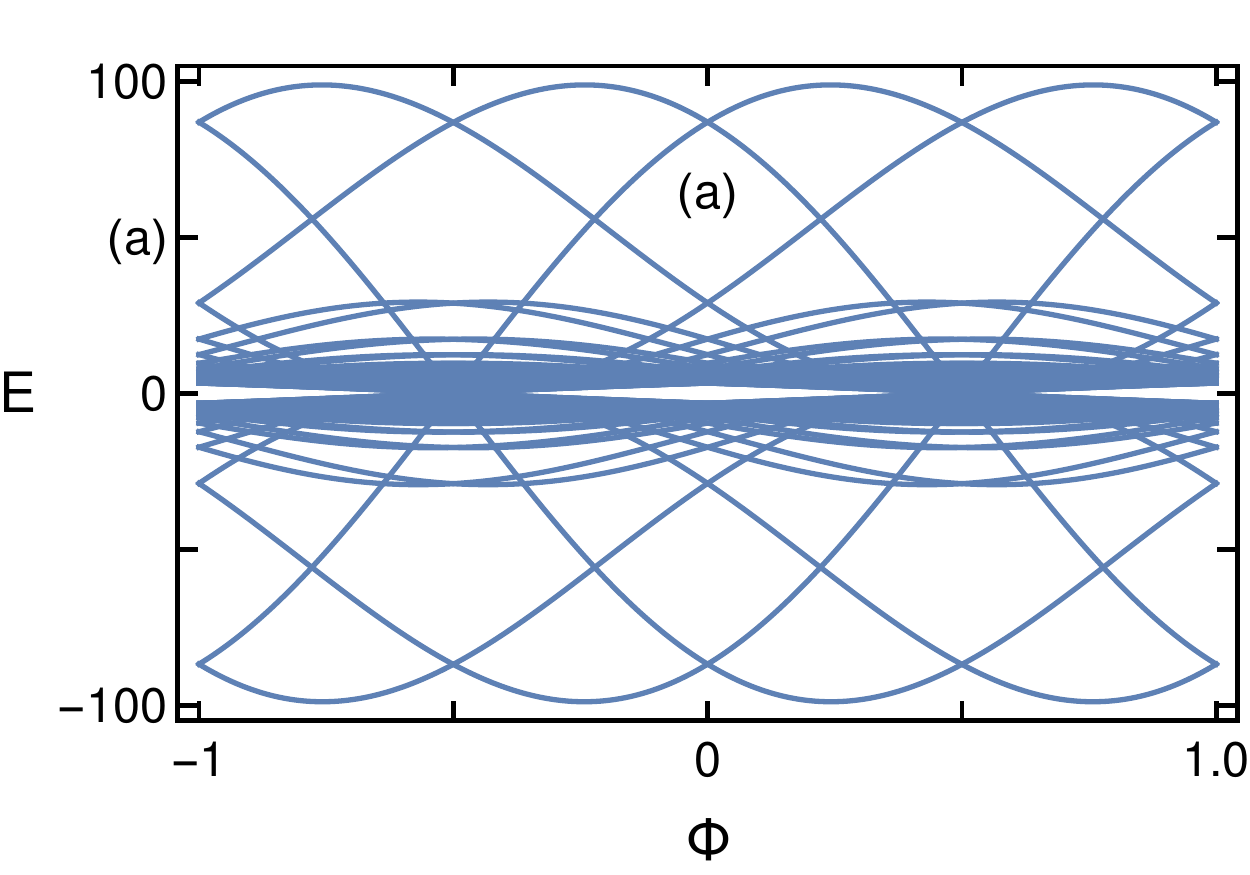}
\includegraphics[height=4.5cm,width=0.40\linewidth]{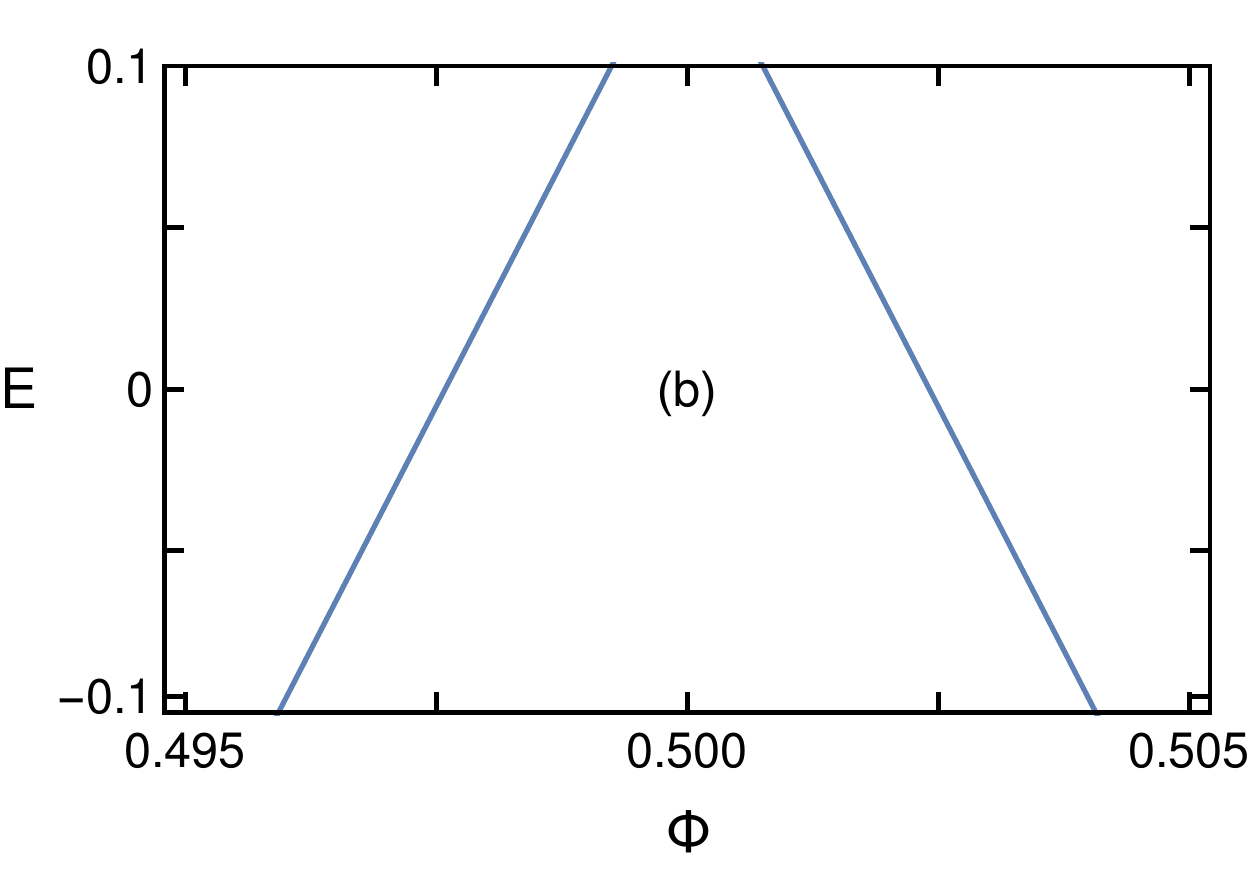}
\includegraphics[height=4.5cm,width=0.55\linewidth]{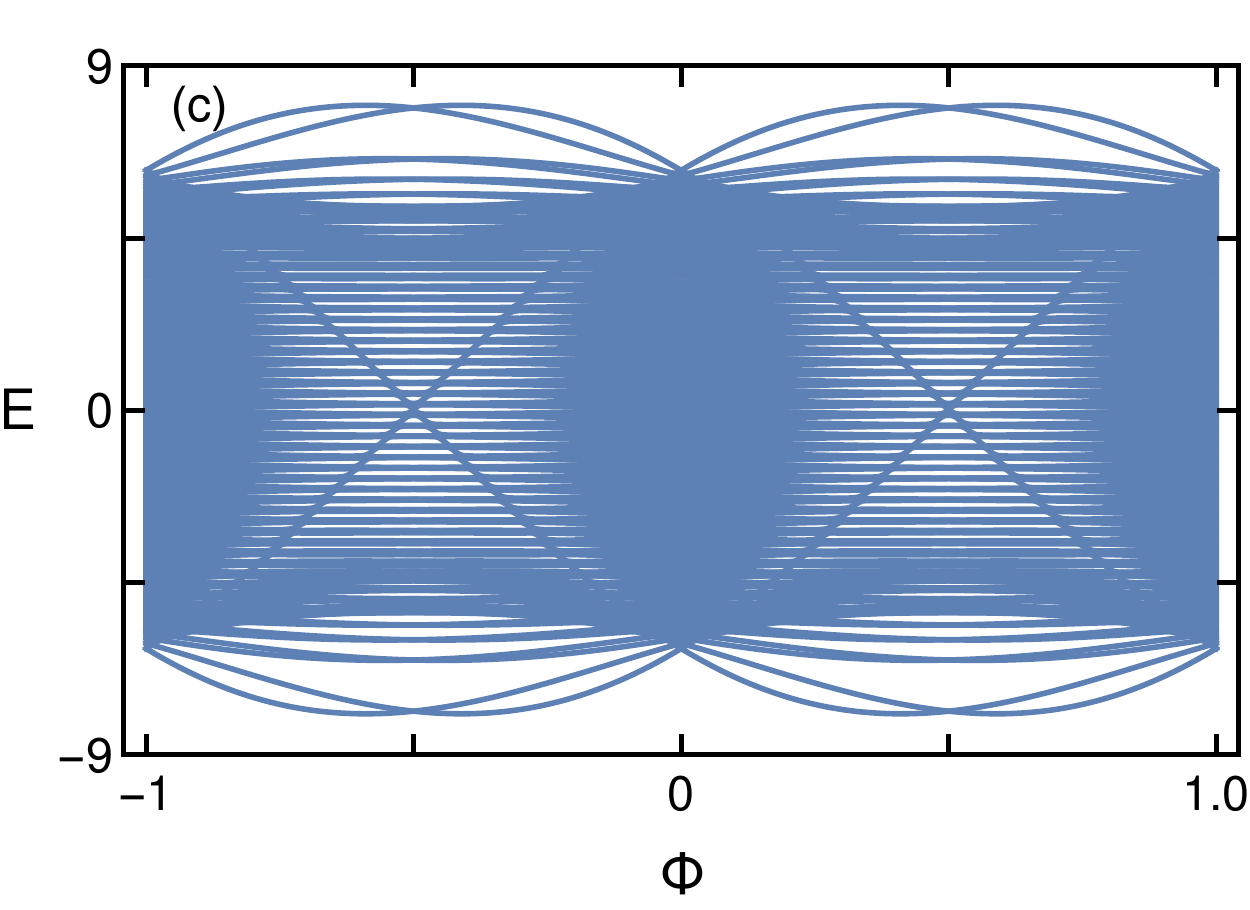}
\includegraphics[height=4.5cm,width=0.39\linewidth]{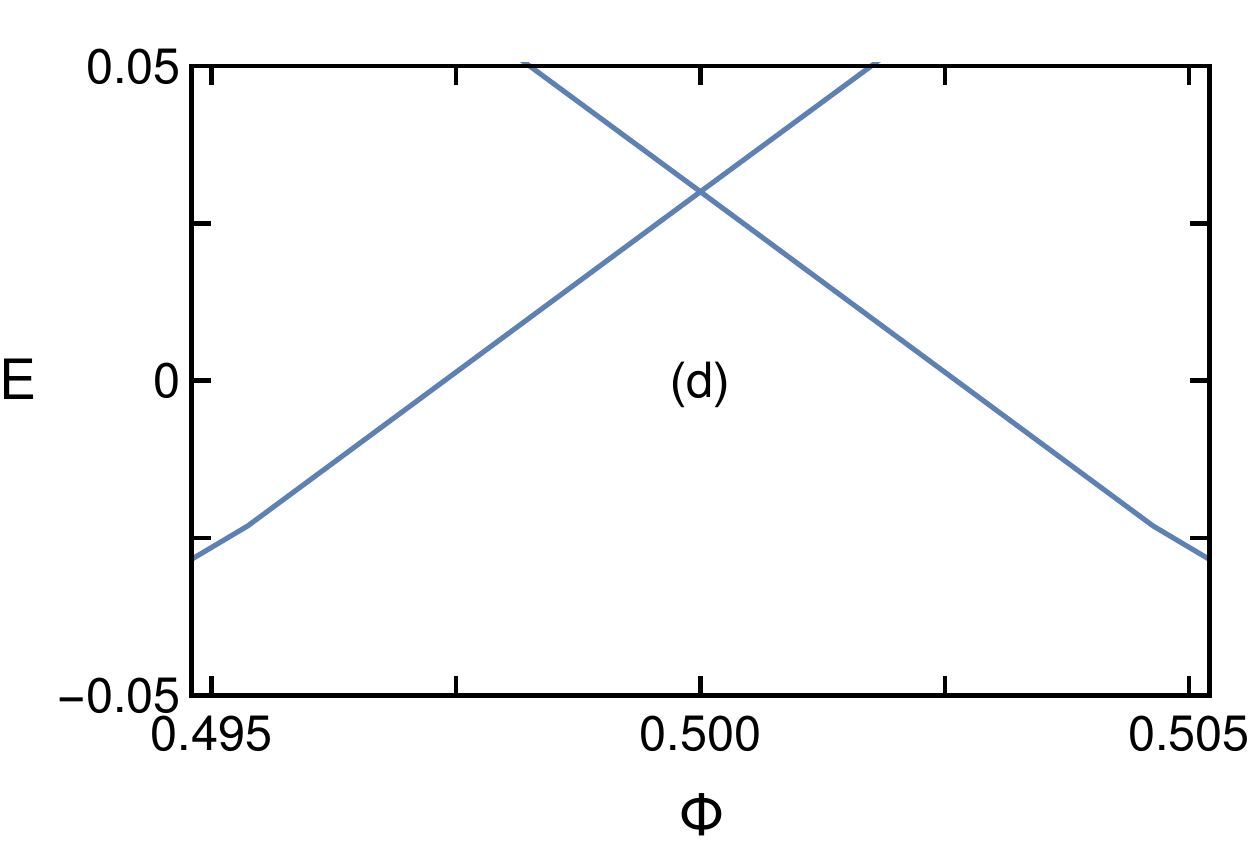}
\caption{(Color online) Left panel: Plot of energy spectra ($E$-$\Phi$) for the disordered ring ($N=91$) 
with disorder strength $W=3$ in presence of RSOI $t_{rso}=1.5$ (a) and in absence of RSOI $t_{rso}=0$ (c). Right 
panel: Illustrations of figure (a) and (c) around $E=0$ are presented in figure (b) and (d) respectively.}
\label{spec_Rdisorder}
\end{figure}
%%%%%%%%%%%%%%%%%%%%%%%%%%%%%%%%
choose the on-site potential energies of the ring randomly from a ``Box" distribution function within the range 
($-W/4$,$W/4$) describing the disorder strength $W$. In Fig.~\ref{spec_Rdisorder}, we show the full energy spectrum 
\ie $E$ vs. $\Phi$ both in presence (Fig.~\ref{spec_Rdisorder}(a)) and absence (Fig.~\ref{spec_Rdisorder}(c)) of RSOI. 
Their zoomed-in versions around $\Phi=\Phi_0/2$ are presented in Figs.~\ref{spec_Rdisorder}(b) and (d) respectively. 
The full spectrum of the isolated disordered ring looks very much similar to that obtained in absence of disorder, but 
both in presence and absence of RSOI as compared with Fig.~\ref{spec_rashba}(a) and Fig.~\ref{spec}(a), respectively. 
The disorder strength is taken as $W=3$. Similar to the case of clean system we have also a central gap around $\Phi=0$ 
in presence of RSOI. There is no such zero-energy crossing at $\Phi=0$ in presence of static disorder also. To examine 
the spectra at half-flux quantum values we zoom-in the full spectrum around $\Phi=\Phi_0/2$ (see Fig.~\ref{spec_Rdisorder}(b)) 
%%%%%%%%%%%%%%%%%%%%%%%%%%%%%%%%
\begin{figure}[!thpb]
\centering
\includegraphics[width=1.0\linewidth]{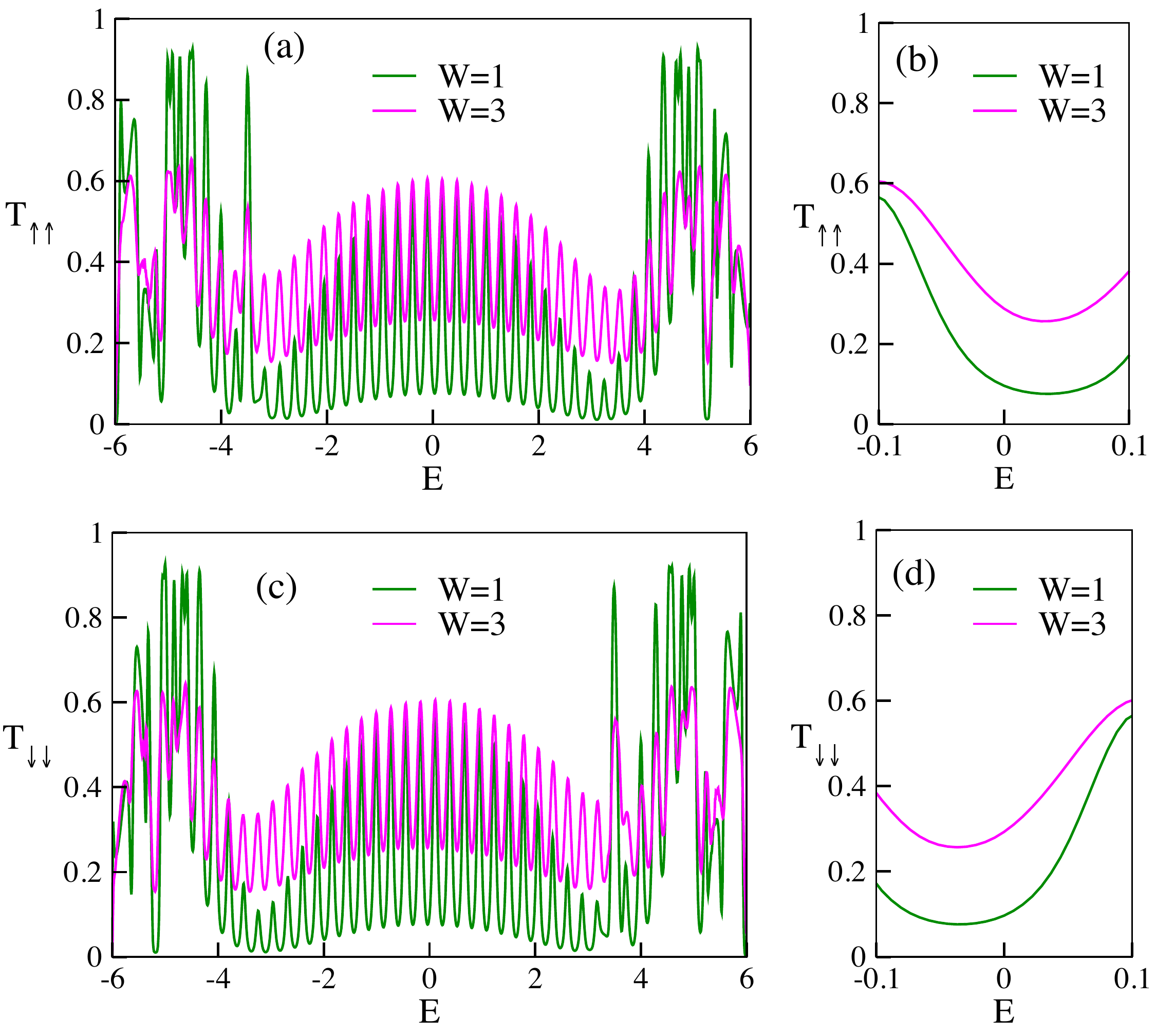}
\caption{(Color online) Behaviors of average transmission probabilities (a) $T_{\uparrow\uparrow}$ and 
(c) $T_{\downarrow\downarrow}$ are displayed as a function of incoming electron energy $E$ for a disordered 
ring ($N=91$) with $\Phi=\Phi_0/4$ in absence of RSOI ($t_{rso}=0$). (b) and (d) represent the zoomed-in 
diagrams of (a) and (c), respectively. The green and magenta colors correspond to the disorder strength 
$W=1$ and $W=3$, respectively. We consider that $60$ sites of the ring are disordered and the result is 
averaged over $100$ disorder configurations.}
\label{te_phi25W}
\end{figure}
%%%%%%%%%%%%%%%%%%%%%%%%%%%%%%%%
and observe the absence of zero energy crossings too. Comparing Fig.~\ref{spec_Rdisorder}(a) with 
Fig.~\ref{spec_rashba}(a), we observe that the difference between the two figures are very small. The difference 
will be much more prominent with higher disorder strength. On the other hand, in absence of RSOI, the zero energy 
crossing gets shifted due to the disorder and the amount of shift depends on the disorder strength (see Fig.~\ref{spec_Rdisorder}(d)). 
That means the zero-energy crossings that we obtain in our helical ring spectra are not immune to disorder. To be 
noted, we consider $60$ out of $91$ sites of the ring as disordered.

In order to examine the transmission characteristics of the electrons in presence of static random disorder we plot 
disorder averaged transmission probabilities, $T_{\uparrow\uparrow}$ and $T_{\downarrow\downarrow}$, in Fig.~\ref{te_phi25W}(a) 
and Fig.~\ref{te_phi25W}(c) and their zoomed-in (around $E=0$) versions in Fig.~\ref{te_phi25W}(b) and Fig.~\ref{te_phi25W}(d)
respectively. Here, green and magenta colors correspond to $W=1$ and $W=3$, respectively. We consider $t_{rso}=0$. 
In this analysis, $2/3$-rd of the total number of sites of the ring are taken as disordered sites and the average is taken over 
$100$ random disorder configurations. We notice that with the enhancement of disorder strength, transmission increases upto a certain 
value of energy and then decreases. Now, in ordinary periodic ring with nearest-neighbor hopping, transmission 
always decreases as soon as we introduce disorder. The higher the disorder strength the lower is the transmission 
probability. Now, manifestation of Anderson type localization in $1$D disordered systems is very well 
known~\cite{anderson1958absence}. Presence of random site potentials cause localization of the electronic 
eigenstates which are basically extended in absence of disorder. With the enhancement of the disorder strength 
the number of localized states increases. As a result transmission probability reduces. The reduction is more 
pronounced when the disorder strength becomes higher than the hopping integral. Specifically, when $W<<t$, 
the Bloch waves are weakly scattered by the random potential. On the other hand, 
%%%%%%%%%%%%%%%%%%%%%%%%%%%%%%%%
\begin{figure}[!thpb]
\centering
\includegraphics[width=1.0\linewidth]{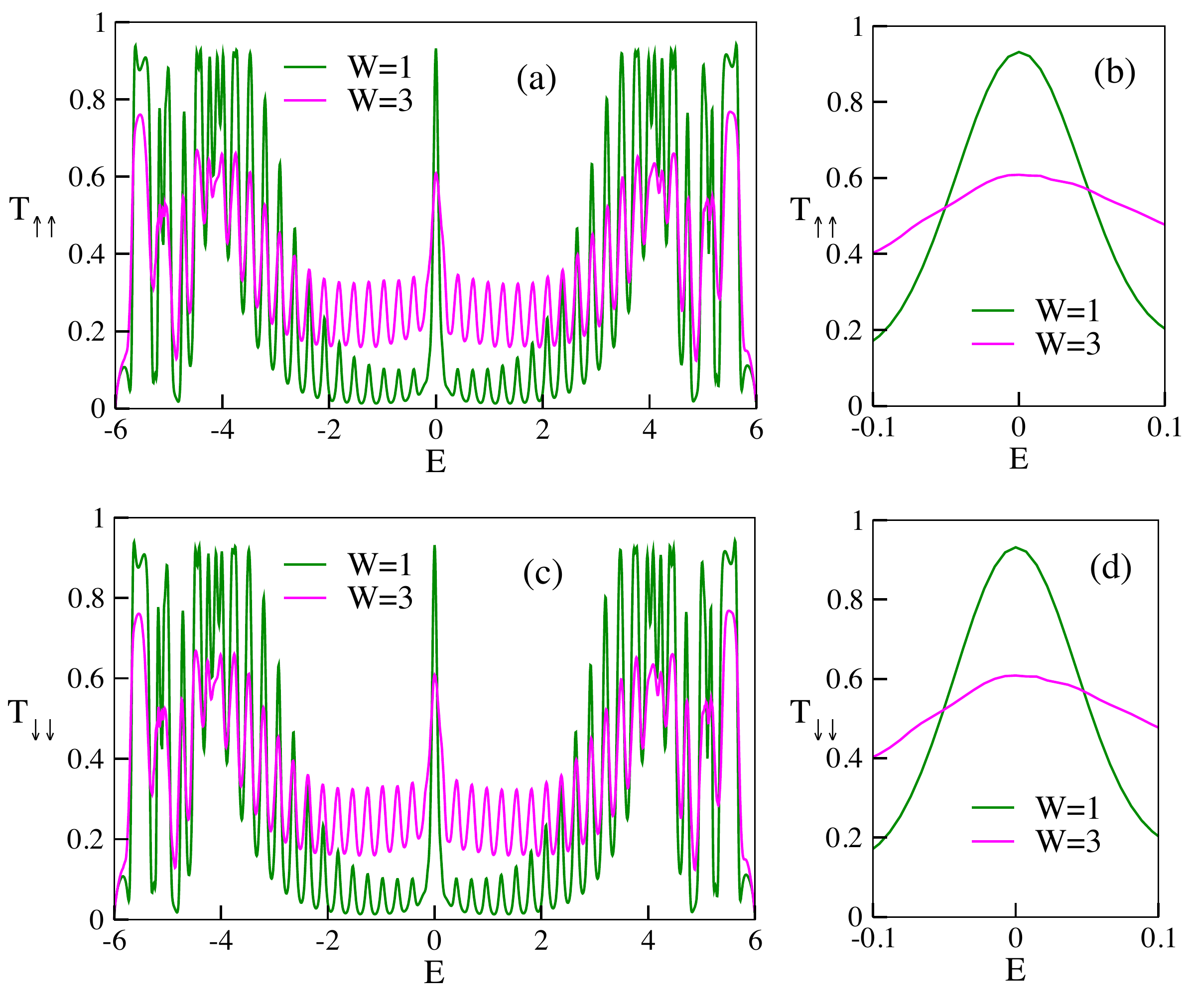}
\caption{(Color online) Plots of transmission probabilities (a) $T_{\uparrow\uparrow}$ and 
(c) $T_{\downarrow\downarrow}$ as a function of incoming electron energy $E$ for a disordered ring are depicted 
for $\Phi=\Phi_0/2$ in absence of RSOI ($t_{rso}=0$). (b) and (d) represent the zoomed-in diagrams of (a) and (c), 
respectively. The green and magenta colors indicate the results corresponding to the disorder strength $W=1$ and 
$W=3$, respectively. Other parameters are fixed to the values same as in Fig.~\ref{te_phi25W}.}
\label{te_phi5W}
\end{figure}
%%%%%%%%%%%%%%%%%%%%%%%%%%%%%%%%
for the condition $W>>t$, the electronic states get localized and they fall off as $e^{-r/\xi}$, $\xi$ being the 
localization length. The system starts behaving like an insulator~\cite{soukoulis1999electronic}.

In contrast to the short-range hopping, when we consider long-range hopping localization phenomenon can not dominate 
the electrons to transmit extendedly due to the presence of higher order hoppings. Long-range hopping basically induces 
an infinite number of resonances~\cite{celardo2016shielding}. As a consequence localization cannot become so effective. 
Even extended state can be obtained in presence of long-range correlated disordered system~\cite{zhang2002localization,
cheraghchi2005localization}. 
From Fig.~\ref{te_phi25W}, 
we observe that with the increase of the disorder strength $W$ the transmission increases upto a certain energy value. 
Then it decreases with the further increase of disorder strength. We cannot predetermine the behavior of the transmission 
amplitudes. Also, we see that there is no peak at zero-energy value similar to the case of clean system as presented in 
Fig.~\ref{te_phi}(e). If we further increase the disorder strength then transmission decreases for all the energy 
values due to localization of the electronic wave-functions. 
 
We also study the effect of disorder on the electron transmission in helical ring with long-range hopping for 
other finite values of the AB flux. For illustration, when we tune the magnetic flux from $\Phi_0/4$ to $\Phi_0/2$ 
we get similar effect of disorder on the transmission probability. The latter means in presence of long-range hopping 
whatever may be the value of magnetic flux we cannot predict whether the transmission through the disordered 
ring would decrease or increase unless we apply very strong disorder. In Figs.~\ref{te_phi5W}(a) and (c) we 
present $T_{\uparrow\uparrow}$ and $T_{\downarrow\downarrow}$,
%%%%%%%%%%%%%%%%%%%%%%%%%%%%%%%%
\begin{figure}[!thpb]
\centering
\includegraphics[width=1.0\linewidth]{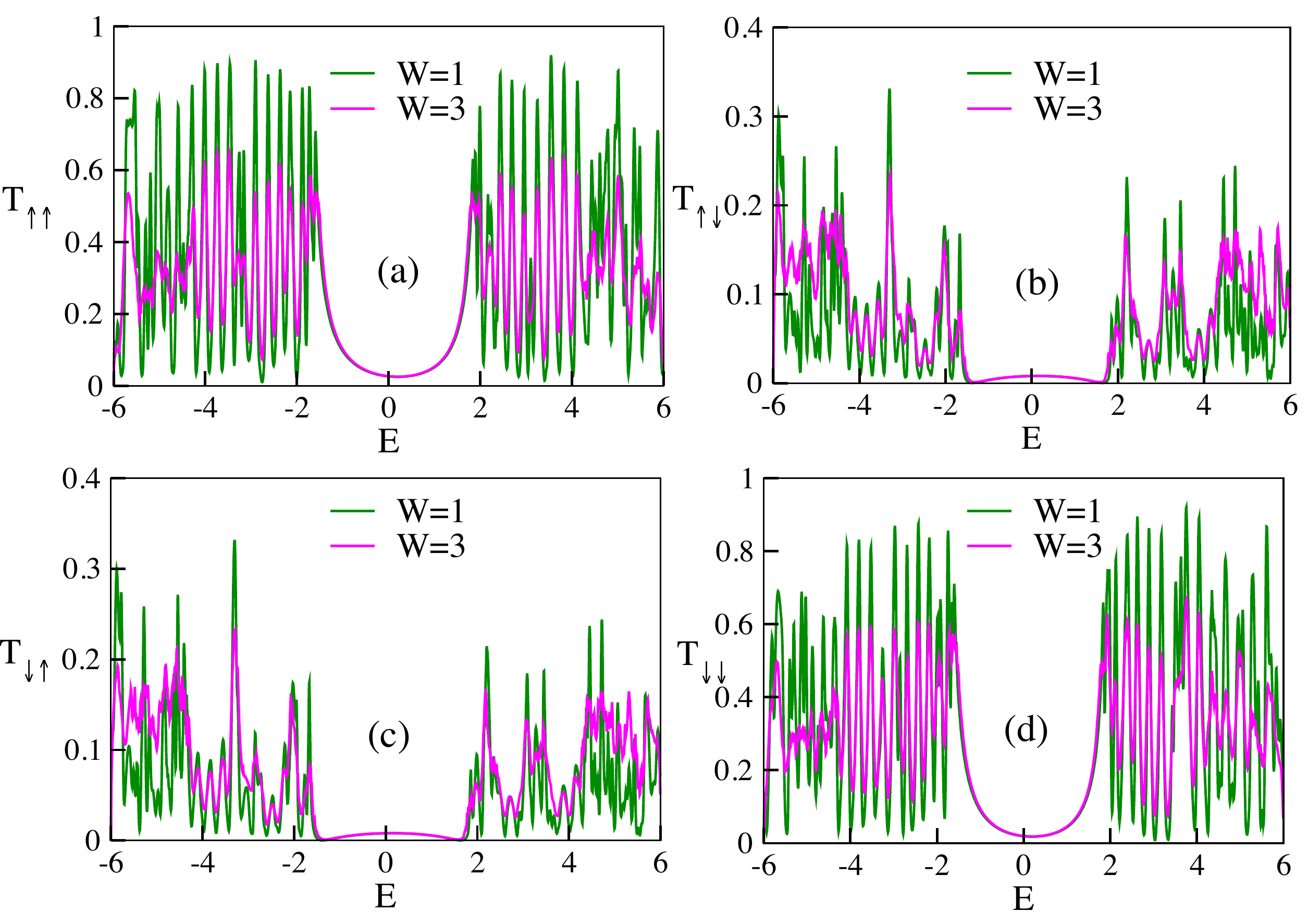}
\caption{(Color online) Quantum-mechanical transmission probabilities (a) $T_{\uparrow\uparrow}$, 
(b) $T_{\uparrow\downarrow}$, (c) $T_{\downarrow\uparrow}$ and (d) $T_{\downarrow\downarrow}$ vs. incoming electron 
energy $E$ are shown for a disordered ring ($N=91$) in presence of RSOI ($t_{rso}=1.5$) and AB flux ($\Phi=\Phi_0/4$). 
The green and magenta colors correspond to disorder strength $W=1$ and $W=3$, respectively. Rest of the parameter 
values are kept same as in Fig.~\ref{te_phi25W}.}
\label{te_Rphi25W}
\end{figure}
%%%%%%%%%%%%%%%%%%%%%%%%%%%%%%%%
respectively. Whereas Figs.~\ref{te_phi5W}(b) and (d) correspond to their zoomed-in versions around $E=0$. 
To be noted, the zero-energy peak for $\Phi=\Phi_0/2$ remains as it was when $W<t$. The peak height decreases with 
the increase of disorder strength. Also, the peak positions get shifted with the rise of the disorder strength. In 
this figure, we get very slight movement of the peak. If we further increase $W$ the shift will be much more 
prominent. Moreover, the plots of $T_{\uparrow\uparrow}$ and $T_{\downarrow\downarrow}$ are exactly similar to each 
other like the phenomenon happened in absence of disorder for $\Phi=\Phi_0/2$.

So far, RSOI has not been taken into consideration in this sub-section while calculating the 
transmission probabilities in presence of disorder. In Fig.~\ref{te_Rphi25W} we show the results in presence of RSOI ($t_{rso}=1.5$) and 
AB flux $\Phi=\Phi_0/4$. Figs.~\ref{te_Rphi25W}[(a), (b), (c) and (d)] correspond to $T_{\uparrow\uparrow}$, $T_{\uparrow\downarrow}$,  
$T_{\downarrow\uparrow}$ and $T_{\downarrow\downarrow}$, respectively. The green and magenta colors have the 
same meaning as in the previous two figures. In presence of RSOI, we have finite spin-flip transmission 
probabilities for both $T_{\uparrow\downarrow}$ and $T_{\downarrow\uparrow}$ due to the finite possibility of 
spin-flipping of electrons. Now, from Fig.~\ref{te_Rphi25W} we notice that in the disordered helical ring when 
we increase the disorder strength all the four transmission probabilities decrease in presence of AB flux. This 
phenomenon is very much similar to the case of periodic ring with only nearest-neighbor hopping~\cite{cheung1988persistent}. 
From literature, we know that RSOI itself may give rise to localization of electronic wave functions. This localization 
phenomenon can be manifested as the spin-precession due to Rashba field alone~\cite{bercioux2004rashba}. 
The overall effect of AB flux and RSOI is the reduction of the transmission of electrons. To be mentioned, 
%%%%%%%%%%%%%%%%%%%%%%%%%%%%%%%%
\begin{figure}[!thpb]
\centering
\includegraphics[width=1.01\linewidth]{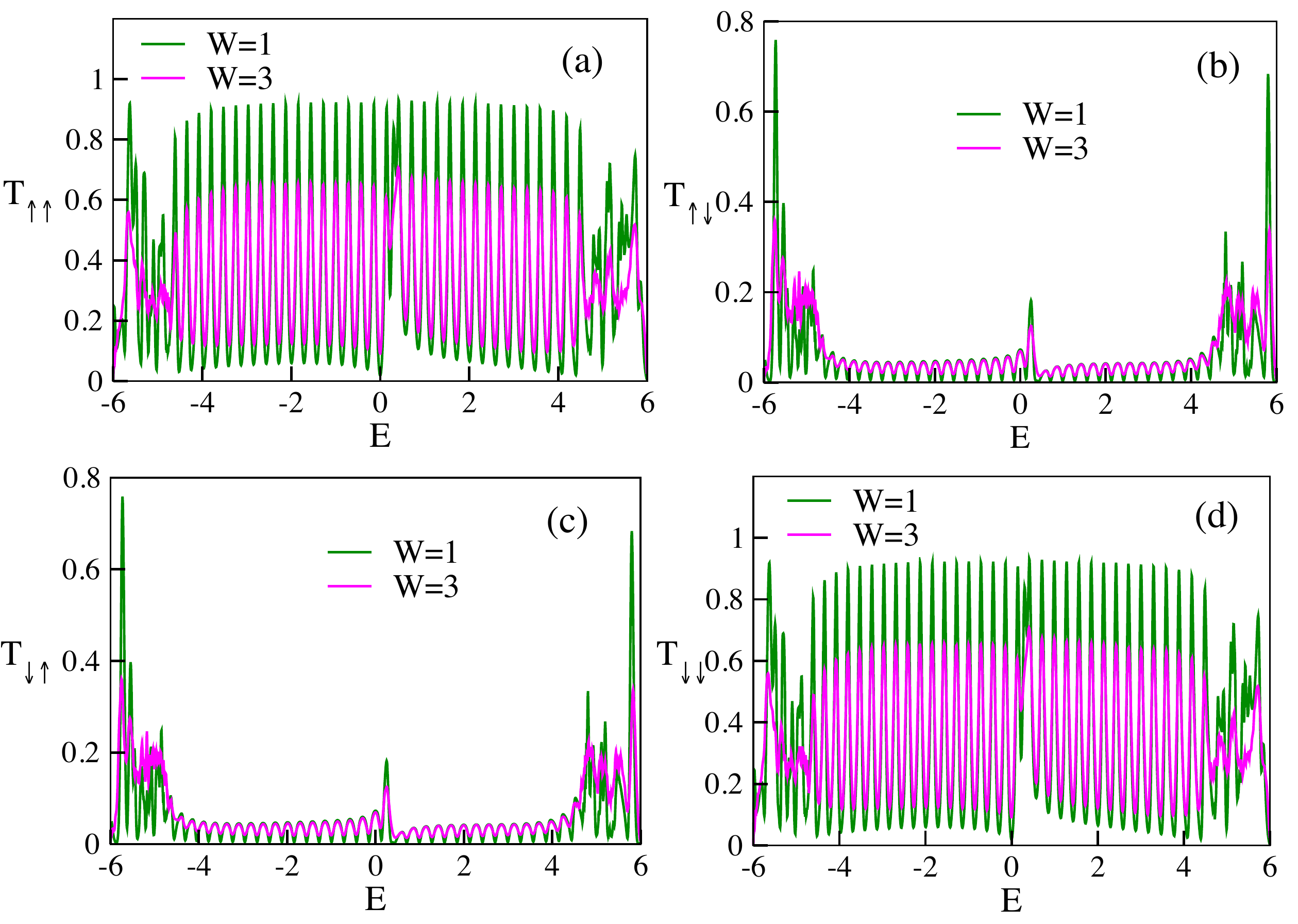}
\caption{(Color online) Behavior of Transmission probabilities (a) $T_{\uparrow\uparrow}$, 
(b) $T_{\uparrow\downarrow}$, (c) $T_{\downarrow\uparrow}$ and (d) $T_{\downarrow\downarrow}$ are shown as a 
function of incoming electron energy $E$ for the same disordered ring as taken in Fig.~\ref{te_Rphi25W} but with 
AB flux ($\Phi=\Phi_0/2$). The green and magenta colors represent the results for the  disorder strength $W=1$ 
and $W=3$, respectively. Other conditions are kept same as in Fig.~\ref{te_phi25W}.}
\label{te_Rphi5W}
\end{figure}
%%%%%%%%%%%%%%%%%%%%%%%%%%%%%%%%
in presence of static disorder the central gap appearing in the spectra is robust to disorder. Gap-widths 
do not change with the increase of disorder strength.

Also, we check our result for $\Phi=\Phi_0/2$ which is illustrated in Fig.~\ref{te_Rphi5W}. For this value of AB flux, transmission 
probabilities decrease similar to the case of $\Phi=\Phi_0/4$. For $\Phi=\Phi_0/2$, the gap disappears as in the case of absence of 
disorder. 

Therefore, we can say that the zero-energy crossings and the zero-energy transmission peaks at $\Phi=0$ and 
half-flux quantum values are not robust to non-magnetic impurity as evident from the energy spectra as well 
as the transmission spectra of the helical ring in presence of RSOI. 

At the end, we would like to emphasize that in order to study the effect of disorder on the transport properties of the helical ring 
we have chosen $2/3$-rd sites of the ring as disordered. Our results are valid for other disorder configurations even 
in presence of single impurity. To visualize the prominent effect of disorder we incorporate more than single impurity as 
in our model we consider only a single propagating mode or channel~\cite{deo1993persistent,chowdhury2008large,bandopadhyay2004hartman}.

%---------------------------------
\section{Summary and Conclusions}
\label{concl}
%---------------------------------

To summarize, we have explored the spin-dependent transmission phenomena in a $1$D ring with long-range hopping. 
In order to introduce the time-reversal counter part, we have considered the signs of the hopping corresponding 
to the two opposite spins are different. The functional form of the hopping has allowed us to manage the 
periodicity very nicely. We have applied AB flux along the axis of the ring. Our model is described in 
tight-binding framework. Using Green's function technique, we have calculated the transmission probabilities of 
the electrons through the ring and show the possibility of getting finite spin polarization by tuning 
AB flux only. The polarization also depends on the energy of the incoming electron.

Two counter-propagating states with opposite spins called as helical states carry the current throughout the ring. 
These helical states are very much similar to the $1$D edge states of a $2$D topological insulator. They are 
characterized by linear dispersion relations. Similar crossings are also present in our helical ring model energy 
spectra. In topological insulator Rashba spin-orbit interaction plays an important role. Also, the helical edge 
states are immune to non-magnetic impurity. To verify whether the counter-propagating edge states obtained in our 
case are protected by topology or not we have investigated the electron transport properties in presence of Rashba 
spin-orbit interaction and static disorder. The helical states obtained in our model are sensitive to individual as 
well as combined effects of RSOI and disorder. RSOI destroys the zero-energy state and a gap appears around $E=0$. 
Whereas, presence of static disorder within the system results in shifting of those zero-energy state as well as the 
reduction in magnitude of the zero-energy transmission peak. Indeed, they also get affected in presence of both RSOI 
and disorder. Therefore, we conclude that the two counter-propagating states of the ring do not mimic the topological 
insulator edge states as the zero-energy transmission peaks as well as zero-energy crossings in the spectra are not 
robust to static random disorder unlike the edge states of topological insulator. 

Finally, we have done a model calculation with some parameter values. For example, we have taken the hopping parameter 
within the ring as $t=2$ eV for bare hopping integral and for Rashba $t_{rso}=1.5$ eV. Our result is valid as long as 
$t_{rso}<t$. For this Rashba value we also do not need to worry about the Rashba decay length towards the inside of the 
ring (tangential direction). It will not affect our main result depending on the aspect ratio (width/radius) of the 
ring~\cite{dfrustaglia}. We have chosen $\tau$ in order to study the strong coupling regime between 
the ring and the leads. 
One can also concentrate on the weak-coupling regime for exploring the transmission probability. In that case, our main 
result will remain invariant, only the peak widths will change. With the change of the other parameter values say $t$ or 
$t_0$ our results will change quantitatively keeping the qualitative nature consistent.

As far as the practical realization of our model is concerned, a quantum ring may be fabricated at the interface of two 
semiconducting materials possessing significant RSOI. For instance, in InAs semiconductor, the Rashba parameter 
$\alpha \sim 2 \times 10^{-11}$ eV m~\cite{mourik2012signatures}. The strength of hopping due to Rashba is related to the 
Rashba parameter as, $\alpha /(2 \delta)$ where $\delta$ is the lattice constant~\cite{moca2005longitudinal}. If we take 
$\delta \sim 5 $ nm then $t_{rso}\sim 0.2$ eV whereas for the above-mentioned lattice constant the nearest neighbor 
hopping integral is $\sim 2$ eV~\cite{mireles2001ballistic}. The magnitude of the external magnetic field can be 
$B\sim 3.2~\rm mT$ for a ring of radius $r\sim 0.2~\rm \mu m$~\cite{moca2005longitudinal}. For our analysis, we have taken 
the parameter values particularly $t_{rso}$ higher in magnitude. However for all other values of Rashba hopping 
strength, the quantitative values of our results will change keeping the qualitative nature unchanged as long as $t_{rso}<t$.

%----------------------------
%\section{acknowledgements}
%----------------------------
\acknowledgments{PD would like to acknowledge Shumpei Masuda, Florian Gebhard and Diptiman Sen for useful discussions and 
comments. PD thanks Science and Engineering Research Board (SERB), Department of Science and Technology (DST), India for 
the partial financial support through National Post-Doctoral Fellowship (File No. PDF/2016/001178). AMJ also thanks DST, 
India for financial support. Finally, we would like to acknowledge Indian citizens for supporting research in basic science.}

%\end{acknowledgements}
%----------------------------

\appendix*
\section{Analytical treatment of the dispersion relation with the AB flux}
\label{appendix}

The energy dispersion relation for the helical ring can be obtained analytically as follows. 

Let us first consider the situation where there is no AB flux and Rashba spin-orbit interaction. For this simplest 
case we have no off-diagonal term in the hopping matrix. We can decouple matrix for the two spin states and derive 
the relation for each spin of the electron separately. The long-range hopping between $m$-th and $n$-th site in 
absence of AB flux and Rashba spin-orbit interaction is given by (for a particular $\sigma$ say +1),
\beq
t_{m,n}= i t  (-1)^{m-n} \frac{1}{(N/\pi)\sin{[\pi(m-n)/N]}}.
\eeq
Now, we start from the Schr\"{o}dinger equation and obtain the following difference equation,
\beq
E \psi_n=-it \sum \limits_{m \ne n} (-1)^{m-n} \frac{1}{(N/\pi)\sin{[\pi (m-n)/N]}} \psi_m.
\eeq
As the ring is periodic, we take Bloch wave solution ($\psi_m \sim e^{i k ma}$) and get
\beq
E=it\sum \limits_{m \ne n}(-1)^{m-n} \frac{1}{(N/\pi)\sin{[\pi (m-n)/N]}} e^{i k (m-n)a} .
\label{e1}
\eeq
Applying the periodic boundary condition, we have 
\beq
k=\frac{2 \pi l}{N a} 
\eeq
where $l$ is an integer and it runs within the range 

\beq
-\frac{(N-1)}{2}\leq l <\frac{(N-1)}{2}.
\eeq

Simplifying Eq.~(\ref{e1}) we get the dispersion relation as,
\beq
E=t k a.
\eeq
Similarly we can find the $E-k$ relation for the other spin state $\sigma=-1$ for which we get a similar equation 
with a negative sign. Therefore, for any $\sigma$ it is given by,
\beq
E=\sigma t k a
\eeq
where $\sigma=\pm 1$. This linear dispersion relation was primarily obtained 
Refs.~\onlinecite{gebhard1992exact,masuda2012interference}. It is completely in contrast to that obtained in ordinary 
ring with nearest-neighbor hopping where we get cosine function in the dispersion relation~\cite{cheung1988persistent}. 
The linear function is discontinuous at the Brillouin zone boundary keeping the periodicity of $2\pi/a$. This allows 
us to get rid of the Fermion doubling problem which generally arises in other discrete models of helical edge states 
having continuous dispersion relation~\cite{cheung1988persistent}. For continuous function we have two end points of 
the Brillouin zone that are identical to each other. 

This linear dispersion relation of the helical ring also holds in presence of magnetic field. In presence of AB
flux we have,

\bea
E=\sigma t \left(k a + \frac{2 \pi\Phi}{N \Phi_0} \right).
\eea

Here, a question may arise. The energy spectra we have shown in this manuscript is more complicated instead of 
linear behavior. Our numerical spectra exactly match with that obtained by analytical calculation without any 
simplification. Without simplification we have the energy dispersion relation as,

\bea
E=\sum \limits_{m=1}^{(N-1)/2} (-1)^m \frac{2 t \sin\Big[m \Big(ka + \frac{2 \pi \Phi}{N\Phi_0}\Big)\Big]}{(N/\pi) \sin(m \pi/N)}\ .
\label{eksum}
\eea

Additionally, in presence of Rashba spin-orbit interaction we have additional off-diagonal terms in the hopping 
matrices. We have matrix form of the difference equation,
\beq
E \bm{I} \psi_m=
\sum_n \left(\begin{array}{c c}
t_{m,n}^l & -i t_{rso} e^{-i \phi_{m,n}}\\
i t_{rso} e^{i \phi_{m,n}} & - t_{m,n}^l 
\end{array}\right) \psi_n.
\eeq
Now again we assume the Bloch wave form for the wave function and solve it. Finally we arrive at the expression,
\beq
E=\pm \sqrt{\xi(k)^2+\xi^{\prime}(k)^2}\ .
\label{ekRashba}
\eeq
where the form of $\xi(k)$ is exactly same as written in the right hand side of Eq.~(\ref{eksum}) and that for 
$\xi^{\prime}(k)$ is given by,

\beq
\xi^{\prime}(k)=2 t_{rso} \sum_{m=1}^{(N-1)/2} \sin\left[m\left(\pi+ka-\frac{2\pi \Phi}{N\Phi_0}\right)\right].
\eeq

We show Eq.~(\ref{ekRashba}) which is exactly equivalent to the numerical one.

%------Ref--------
\bibliography{bibfile}{}

\end{document}